\documentclass[aip,apl,amsmath,amssymb,reprint,superscriptaddress]{revtex4-1}

\usepackage{mathrsfs,amsfonts} 
\usepackage{bbm}
\usepackage[colorlinks,citecolor=blue,linkcolor=red,urlcolor=blue]{hyperref}
\usepackage{color}
\usepackage{graphicx}
\usepackage{subfigure}
\usepackage{verbatim}
\usepackage{fourier}

\begin{document}

\title{Effect of graphene grating coating on near-field radiative heat transfer}

\author{Minggang Luo}
\email[]{minggang.luo@umontpellier.fr}
\affiliation{Laboratoire Charles Coulomb (L2C), UMR 5221 CNRS-Universit\'e de Montpellier, F-34095 Montpellier, France}

\author{Youssef Jeyar}
\affiliation{Laboratoire Charles Coulomb (L2C), UMR 5221 CNRS-Universit\'e de Montpellier, F-34095 Montpellier, France}

\author{Brahim Guizal}
\affiliation{Laboratoire Charles Coulomb (L2C), UMR 5221 CNRS-Universit\'e de Montpellier, F-34095 Montpellier, France}

\author{Junming Zhao}
\affiliation{School of Energy Science and Engineering, Harbin Institute of Technology, 92 West Street, Harbin 150001, China}
\affiliation{Key Laboratory of Aerospace Thermophysics, Ministry of Industry and Information Technology, Harbin 150001, China}

\author{Mauro Antezza}
\email[]{mauro.antezza@umontpellier.fr}
\affiliation{Laboratoire Charles Coulomb (L2C), UMR 5221 CNRS-Universit\'e de Montpellier, F-34095 Montpellier, France}
\affiliation{Institut Universitaire de France, 1 rue Descartes, Paris Cedex 05, F-75231, France}


\date{\today}
\begin{abstract}

In this work we analyse the near-field radiative heat transfer (NFRHT) between finite-thickness planar fused silica slabs coated with graphene gratings. We go beyond the effective medium approximation by using an exact Fourier Modal Method (FMM) equipped with specific Local Basis Functions (LBF), and this is needed for realistic experimental analysis. In general, coating a substrate with a full graphene sheet has been shown to decrease the NFRHT at short separations (typically for d<100 nm) {compared to the bare substrates}, where the effective medium approximation consistently overestimates the radiative heat flux, with relative errors exceeding 50\%. We show that, by patterning the graphene sheet into a grating, the topology of the plasmonic graphene mode changes from circular to hyperbolic, allowing to open more channels for the energy transfer between the substrates. We show that, at short separations, the NFRHT between slabs coated with graphene gratings is higher than that between full-graphene-sheet coated slabs and also than that between uncoated ones. We show a significant dependence of the radiative heat transfer on the chemical potential, which can be applied to modulate \textit{in situ} the scattering properties of the graphene grating without any geometric alterations. We also compare the exact calculation with an approximate additive one and show that this approximation performs quite well for low chemical potentials. This work has the potential to unveil new avenues for harnessing non-additive heat transfer effects in graphene-based nanodevices.

\end{abstract}
\maketitle 
Near-field radiative heat transfer (NFRHT) has garnered significant attention in recent times, driven by both fundamental explorations and practical applications. When the separation distance is comparable to or less than the thermal wavelength $\lambda_T=\hbar c/k_BT$, then the radiative heat flux between the two involved bodies can go far beyond the Planckian blackbody limit by several orders of magnitude \cite{Rytov1989,Polder1971}. The NFRHT has been widely investigated for many different geometrical configurations, e.g., plane-plane \cite{Loomis1994,Carminati1999,Shchegrov2000,Volokitin2001}, sphere-sphere (including nanoparticle-nanoparticle) \cite{Narayanaswamy2008,Chapuis2008,Manjavacas2012,Nikbakht2018,Messina2018,DongPrb2018,Zhang2019R,He2019APL,Zhang2019T}, nanoparticle-surface \cite{Chapuis2008plate}, grating-involved nanostructures \cite{Yang2017prl,Zheng2022Materials_EMA,Biehs2011gratings,Kan2019prb,Kan2021ijhmt,Yang2017ijhmt}, of which some have been confirmed by pioneering experimental works \cite{Ottens2011,Lim2015,Watjen2016,Ghashami2018,Yungui2018NC,Yungui2023PRAppl,DeSutter2019,Shen2009,Rousseau2009,Song2015}. In particular, gratings lead to the excitation of high-order diffraction modes that play a relevant role in the NFRHT.
\begin{figure} [tbp]
\centerline {\includegraphics[width=0.5\textwidth]{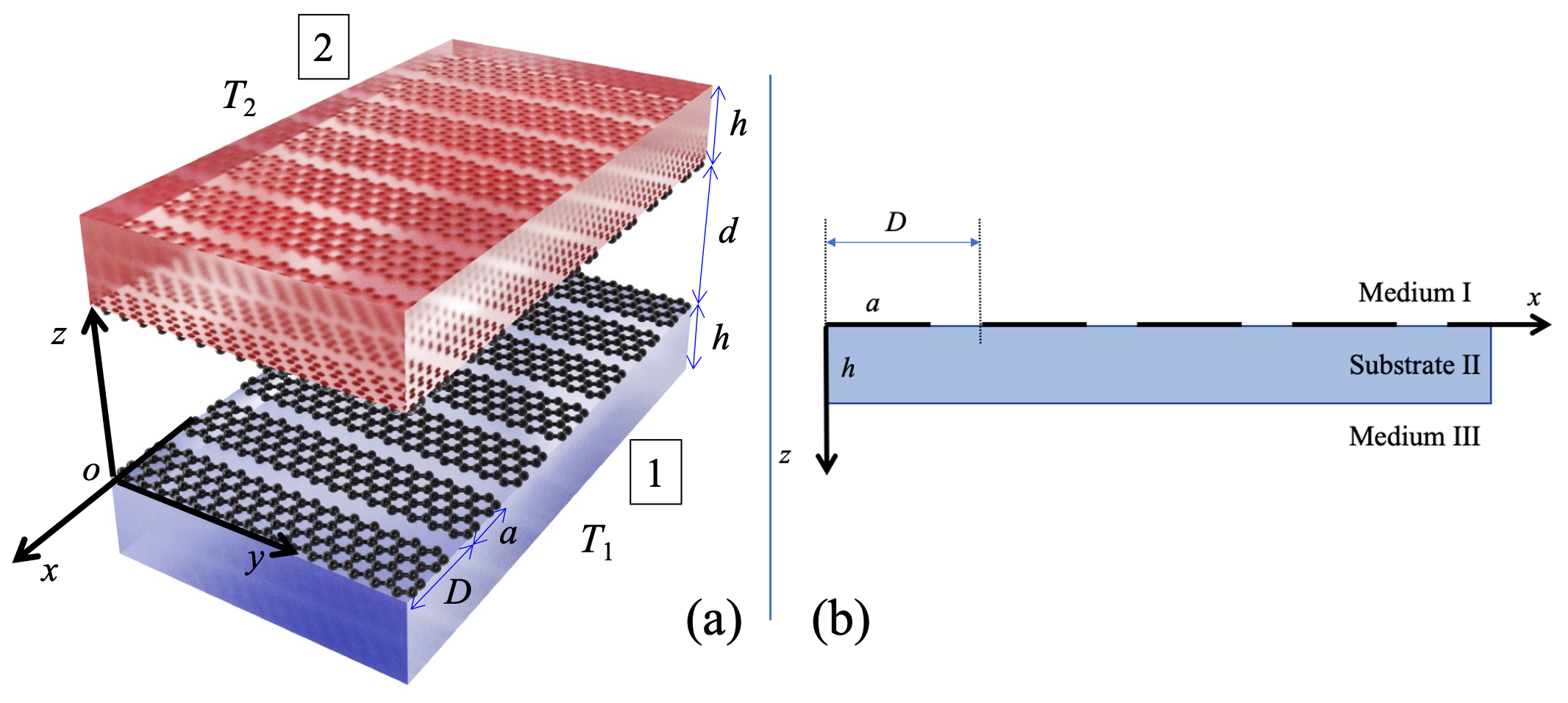}}
     \caption{Sketch of the studied configuration}
   \label{fig:two_gratings_schematic}
\end{figure}

Due to graphene's peculiar optical properties, graphene-based nanostructures have shown novel behaviors in the radiative heat transfer modulation \cite{Lu2022small,Shi2021am,Volokitin2017Dey,Zhao2017prb,He2020ijhmt,He2020ol,Zheng2017,Ognjen2012prb,Svetovoy2012prb,JSWM,GrafBilMA}. In general, at large separations, adding a graphene layer on top of a dielectric slab can dramatically turn a poor heat emitter/absorber into a good one \cite{Svetovoy2012prb}. However, the graphene coating is shown to decrease the NFRHT at short separations (typically for d<100 nm) \cite{Svetovoy2012prb,LiuES2022}. By patterning a suspended graphene sheet into strips (one-dimensional grating) \cite{Liu2015APL} and square arrays (two-dimensional grating) \cite{Yang2020prAppl}, more channels will be opened to allow for radiative energy transfer, where an exact method based on the rigorous coupled-wave analysis (RCWA) is applied to treat the scattering from the grating. When there is a supporting substrate, it is interesting to know whether such a patterning method still works or not and it is this question we are addressing in this work. {More importantly, the supported graphene gratings are more realistic and experimentally relevant compared to the suspended ones in the literature.} In our numerical investigations, we will use another exact method rather than the aforementioned RCWA, \textit{i.e.,} the Fourier Modal Method with Local Basis Functions (FMM-LBF) which is far more efficient for {two-dimensional} strip gratings \cite{Guizal, Taiwan_LBF, PRE23} {(for more detailed discussion of the two methods please refer to our recent work \cite{Luo2023Casimir_gg})}. It's also worthwhile mentioning that recently NFRHT between graphene strips on top of uniaxial hyperbolic substrates was investigated by using the effective medium theory (EMT) \cite{Zhou2022langmuir}. The applicability of the latter approach {is valid when the separation distance is much larger than the grating period \cite{Liu2014JHT,Biehs2011OE}}. By comparing the exact method to the EMT we will be able to study the EMT approximation validity in graphene-grating coated systems.

In this letter, we use the FMM-LBF to investigate the near-field radiative heat transfer between planar fused silica slabs (see Fig.~\ref{fig:two_gratings_schematic} (a)) coated with graphene gratings. We show that a graphene grating coating a substrate can significantly enhance the {near-field radiative} heat transfer, we also investigate the limits of the the effective medium approximation and introduce and study a second simple {additive} approximation.\\

The physical system under consideration is shown in Fig.~\ref{fig:two_gratings_schematic} (a). We investigate the near-field radiative heat transfer between bodies 1 and 2 that consist of dielectric slabs having the same thickness $h$ and coated with graphene strip gratings separated by a distance $d$. The graphene grating period is $D$ while the width of one single graphene strip is $a$, thus leading to a filling fraction $f=a/D$. Bodies 1 and 2 are maintained at fixed temperatures $T_1$ and $T_2$ respectively, and the environment is maintained at the same temperature as body 1 (\textit{i.e.}, $T_1$). The net power flux $\varphi$ received by body 1 (energy per unit surface and time) can be defined as \cite{Messina2014,Messina2011PRA} 
\begin{equation}
\varphi= \sum_p \int \frac{{\rm d}^2 {\rm \textbf{k}}}{(2\pi)^2} \int_0^{+\infty} \frac{{\rm d} \omega}{2\pi}<p,{\rm \textbf{k}}|\mathcal{O}|p,{\rm \textbf{k}}>,
\label{HF}
\end{equation}
where $p$ is the polarization index, $p=$ 1, 2 correspond to TE and TM modes, $\omega$ is the angular frequency, ${\rm \textbf{k}}=(k_x, k_y)$, $k_x$ and $k_y$ are the parallel components of the wave vectors in the standard $(x, y, z)$ coordinates, the operator $ \mathcal{O}$ is defined in the TE/TM basis as \cite{Messina2014} 
\begin{widetext}
\begin{equation}
\mathcal{O}= \hbar \omega n_{21}  U^{(2,1)} \left[  f_{-1}(\mathcal{R}^{(2)-}) - \mathcal{T}^{(2)-} \mathcal{P}_{-1}^{({\rm pw})}  \mathcal{T}^{(2)-\dagger} \right] U^{(2,1)\dag } \left[  f_{1}(\mathcal{R}^{(1)+}) - \mathcal{T}^{(1)-\dagger} \mathcal{P}_{1}^{({\rm pw})}  \mathcal{T}^{(1)-} \right]  ,
\label{O_operator}
\end{equation}
\end{widetext}
where $\hbar$ is the reduced Planck constant, $n_{21}=n(\omega,T_{2}) - n(\omega,T_{1})$, $n(\omega,T)=1/( e^{\frac{\hbar \omega}{k_{\rm B} T}}-1 )$, $U^{(2,1)}=(1-\mathcal{R}^{(2)-}\mathcal{R}^{(1)+})^{-1}$, the $\mathcal{R}^{(1)+}$ and $\mathcal{R}^{(2)-}$ ($\mathcal{T}^{(1)-}$ and $\mathcal{T}^{(2)-}$ ) are the reflection operators (transmission operators) of body 1 and body 2 in the TE/TM basis, $\dagger$ is for conjugate operation, the $+$ and $-$ superscripts in the reflection and transmission coefficients correspond to the propagation directions, $
 \left \langle p,{\textbf{k}}|\mathcal{P}_{n}^{\rm (pw/ew)}|p',{\textbf{k}}' \right\rangle 
 =k_{z}^{n}	\left \langle p,{\textbf{k}}|\mathcal{\prod}^{\rm (pw/ew)}|p',{\textbf{k}}' \right\rangle$, $k_{z}=\sqrt{k_0^2-\textbf{k}^2}$, $k_0=\omega/c$, $\mathcal{\prod}^{\rm (pw)}$ ($\mathcal{\prod}^{\rm (ew)}$) is the projector on the propagative (evanescent) waves and the auxiliary function  $f_{\alpha}(\mathcal{R})$ is defined as follows \cite{Messina2014,Messina2011PRA}:
\begin{widetext}
\begin{equation}
f_{\alpha}(\mathcal{R})=\left\{
\begin{array}{rcl}
\mathcal{P}_{-1}^{(\rm pw)} - \mathcal{R} \mathcal{P}_{-1}^{(\rm pw)} \mathcal{R}^{\dagger} + \mathcal{R} \mathcal{P}_{-1}^{(\rm ew)}  -  \mathcal{P}_{-1}^{(\rm ew)} \mathcal{R}^{\dagger}   & & \alpha = -1 \\ \\
\mathcal{P}_{1}^{(\rm pw)} - \mathcal{R}^{\dagger} \mathcal{P}_{1}^{(\rm pw)} \mathcal{R} + \mathcal{R}^{\dagger} \mathcal{P}_{1}^{(\rm ew)}  -  \mathcal{P}_{1}^{(\rm ew)} \mathcal{R}   & & \alpha = 1.
\end{array}
\right.
\label{auxiliary_function}
\end{equation}
\end{widetext}


When calculating the reflection and transmission operators of a single body, we consider the scheme of  Fig.~\ref{fig:two_gratings_schematic}(b). The $\mathcal{R}^-$ and  $\mathcal{T}^-$ are calculated relative to the body with the interface ($z=0$) between medium I and II by using the FMM-LBF (for more details see\cite{Luo2023Casimir_gg}). The size of $\mathcal{R}^-$ and  $\mathcal{T}^-$ is $ 2(2N +1)\times 2(2N +1)$, where $N$ is the truncation order to be chosen large enough to {achieve} convergence. In order to calculate the $\mathcal{R}^{(2)-}$ and  $\mathcal{T}^{(2)-}$ of body 2, we must introduce a phase shift because for this latter the interface lying between medium I and II is now set at $z=d$. Due to this translation, the $\mathcal{R}^{(2)-}$ and  $\mathcal{T}^{(2)-}$ can be modified as follows \cite{Messina2011PRA}:
\begin{equation}
\left\{
\begin{array}{rcl}
\begin{aligned}
 &\left \langle p,{\textbf{k}},n|\mathcal{R}^{(2)-}(\omega)|p',{\textbf{k}}',n' \right\rangle 
 \\&=e^{i\left(k_{z,n}^{}+k_{z,n'}'\right)d}	\left \langle p,{\textbf{k}},n|\mathcal{R}^{-}(\omega)|p',{\textbf{k}}',n' \right\rangle,
\\&\left \langle p,{\textbf{k}},n|\mathcal{T}^{(2)-}(\omega)|p',{\textbf{k}}',n' \right\rangle 
 \\&=e^{i\left(k_{z,n}^{}-k_{z,n'}'\right)d}	\left \langle p,{\textbf{k}},n|\mathcal{T}^{-}(\omega)|p',{\textbf{k}}',n' \right\rangle.
\end{aligned}
\end{array}
\right.
\label{R2_T2_}
\end{equation}


We then focus on the $\mathcal{R}^{(1)+}$ and  $\mathcal{T}^{(1)-}$ of body 1 (whose interface is at $z=0$), which can be obtained from the known $\mathcal{R}^-$ and  $\mathcal{T}^+$ with the following relations\cite{Messina2017PRB}:   
\begin{equation}
\begin{split}
&\left \langle p,{\textbf{k}},n|\mathcal{R}^{(1)+}(\omega)|p',{\textbf{k}}',n' \right \rangle \\ &=  \left\{
\begin{array}{rcl}
\begin{aligned}
&\left \langle p,{\textbf{k}},n|\mathcal{R}^{-}(\omega)|p,{\textbf{k}}',n' \right\rangle & p=p'  \\
&-\left \langle p,{\textbf{k}},n|\mathcal{R}^{-}(\omega)|p',{\textbf{k}}',n' \right\rangle & p\neq p', 
\end{aligned}
\end{array}
\right.
\end{split}
\label{R1+}
\end{equation}
for the reflection operator $\mathcal{R}^{(1)+}$, and 

\begin{equation}
\begin{split}
& \left \langle p,{\textbf{k}},n|\mathcal{T}^{(1)-}(\omega)|p',{\textbf{k}}',n' \right\rangle \\& =\left\{
\begin{array}{rcl}
\begin{aligned}
& \left \langle p,{\textbf{k}},n|\mathcal{T}^{+}(\omega)|p,{\textbf{k}}',n' \right\rangle, p=p' 
\\
&  -\left \langle p,{\textbf{k}},n|\mathcal{T}^{+}(\omega)|p',{\textbf{k}}',n' \right\rangle, p\neq p' ,  
\end{aligned}
\end{array}
\right.
\end{split}
\label{nondiagnoal}
\end{equation}
for the the transmission operator $\mathcal{T}^{(1)-}$.

\begin{figure*} [htbp]
\centerline {\includegraphics[width=0.8\textwidth]{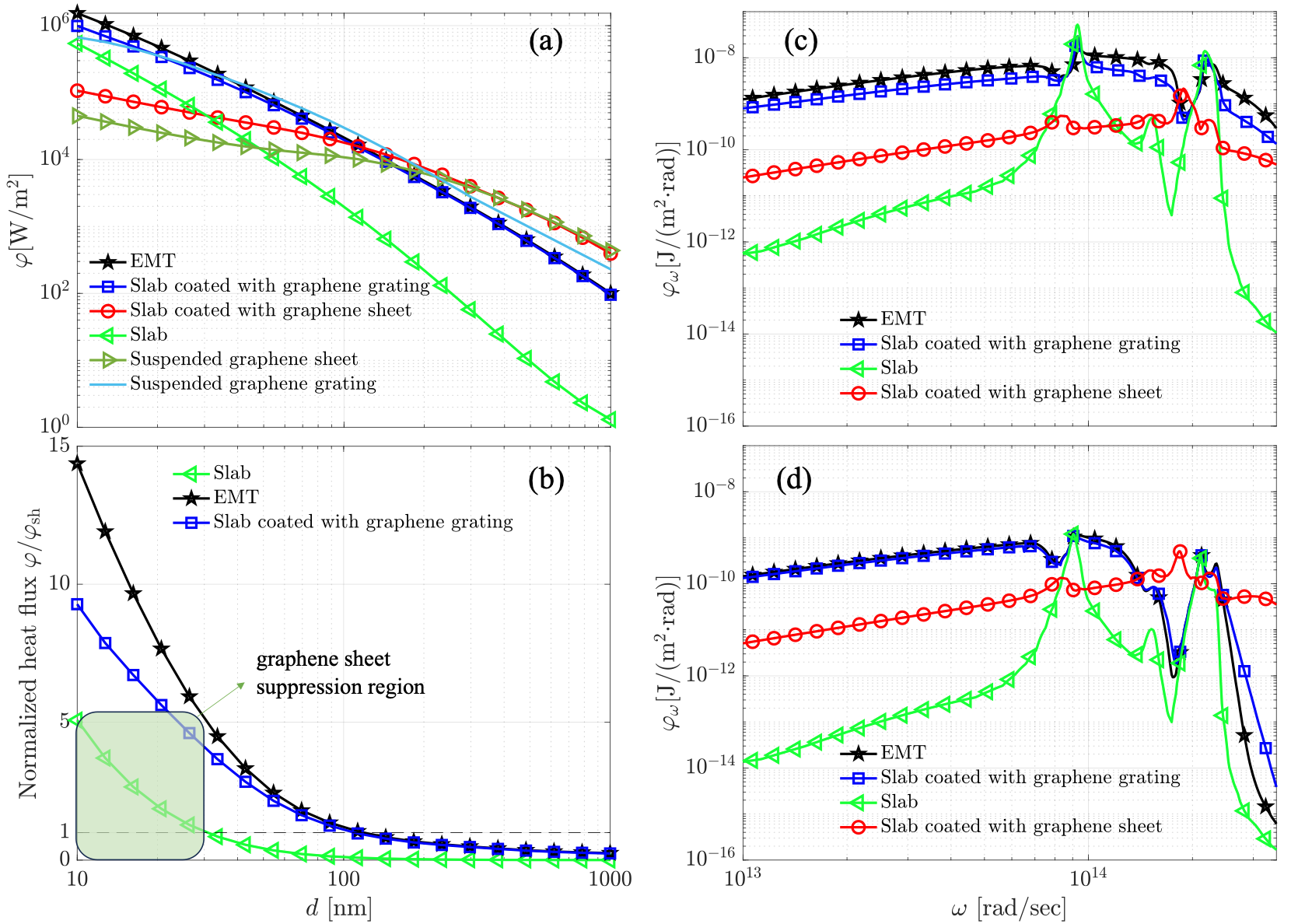}}
     \caption{The dependence of the radiative heat flux on the separation $d$ and the radiative heat flux spectrums at fixed separations: (a) absolute radiative heat flux $\varphi$ between bodies 1 and 2, (b) normalized radiative heat flux $\varphi/\varphi_{\rm sh}$ ($\varphi_{\rm sh}$ is the radiative heat flux between two slabs coated with graphene sheets), (c) radiative heat flux spectrum at $10$ nm, and (d) radiative heat flux spectrum at $50$ nm.}
   \label{fig:distance_dependence}
\end{figure*}

By using the expressions above we can now study the {NFRHT} between the two planar graphene-grating nanostructures shown in Fig.~\ref{fig:two_gratings_schematic} (a). We consider a substrate made of fused silica (SiO$_2$), whose optical data are taken from Ref.\citenum{Palik}. The graphene conductivity, which depends on the temperature $T$ and on chemical potential $\mu$, is the sum of two contributions (an intraband $\sigma_{\textnormal{intra}}$  and an interband $\sigma_{\textnormal{inter}}$ terms) given by\cite{Zhao2017prb,Falkovsky2007,Falkovsky2008,Awan_2016}:
\begin{equation}
\sigma_{\textnormal{intra}} = \frac{i}{\omega + i/\tau} \frac{2 e^2 k_{\rm B} T}{\pi\hbar^2}\ln\left[ 2 \cosh\left(\dfrac{\mu}{2k_{\rm B}T}\right) \right]
\end{equation}
and
\begin{equation}
\sigma_{\textnormal{inter}}= \frac{e^2}{4 \hbar} \left [ G\left(\frac{\hbar \omega}{2}\right) + i \dfrac{ 4\hbar \omega}{\pi}\int_0^{+\infty} \dfrac{G(\xi)-G\left(\hbar \omega /2\right)}{(\hbar \omega)^2 - 4\xi^2} {\rm d}\xi \right],
\end{equation}
where  $e$  is the electron charge, $\tau$ the relaxation time (we use $\tau = 10^{-13}${s}), $G(\xi) = \sinh (\xi/k_B T) / [\cosh (\mu/k_{\rm B} T) + \cosh(\xi/k_{\rm B} T)]$. In Fig.~\ref{fig:distance_dependence} (a), we show the dependence of the radiative heat flux $\varphi$ on the separation $d$. Five cases are considered : (I) planar slabs coated with graphene sheets, (II) planar slabs coated with graphene gratings, (III) bare planar slabs, (IV) suspended graphene sheets, and (V) suspended graphene gratings, respectively. Bodies 1 and 2 are maintained at 290 K and 310 K, respectively.  The radiative heat flux for planar slabs coated with graphene gratings obtained by using the effective medium theory (EMT) is also added for comparison. The parameters for these computations are : $\mu=0.5$ eV, grating period $D=20$ nm, thickness $h=20$ nm and filling fraction $f=0.2$. The truncation order $N$ is fixed at 30 to ensure convergent results.


As shown in Fig.~\ref{fig:distance_dependence} (a), for separations $d>30$ nm, the graphene sheet coating (red circle line) can induce a big enhancement of the radiative heat flux with respect to bare slabs (left green triangle line). However, the graphene sheet cannot enhance the radiative heat flux at short separations ($d<30$ nm), but can bring a significant inhibition, which is consistent with the observation for the SiC substrate/graphene sheet in literature \cite{LiuES2022}. In the ideal case of suspended graphene configurations, it is reported that the patterning of the graphene sheets into ribbons can lead to an enhancement of the radiative heat transfer at short separations due to introduction of hyperbolic graphene plasmons  \cite{Liu2015APL}. Is this enhancement still present when graphene is supported by a substrate, in a much realistic configuration ?  We can now answer this question affirmatively, indeed Fig.~\ref{fig:distance_dependence} (a) shows that the radiative heat flux between the slabs coated with graphene gratings (blue square line) is much larger than that of the bare slabs (light green triangle line). In Fig.~\ref{fig:distance_dependence}(b) we also show the dependence of the normalized radiative heat flux $\varphi/\varphi_{\rm sh}$ on the separation $d$. Here the normalisation factor $\varphi_{\rm sh}$ is the radiative heat flux between two slabs coated with graphene sheets. In the light-green region of Fig.~\ref{fig:distance_dependence}(b), at short separations ($d<30$ nm), we see that NFRHT between bare slabs is much higher than that of the same slabs coated by a graphene sheet. However, when patterning the graphene sheets into gratings, the radiative heat flux can even increase to almost two times that of the bare slabs. Moreover, as can be seen in Figs.~\ref{fig:distance_dependence} (a) and (b), the EMT overestimates the radiative heat flux (relative error to the exact one can be as large as 50$\%$) for short separations ($d<100$ nm).



To understand why the patterning method improves the radiative heat transfer, we show in Figs.~\ref{fig:distance_dependence} (c) and (d) the radiative heat flux spectrum for two different separations, $d=$ 10 nm and 50 nm, respectively. The radiative heat flux spectrum for the bare fused silica slabs has two peaks in the Reststrahlen bands at the two considered separations. This is also the case for the slabs coated with graphene gratings. However, for the slabs coated with graphene sheets at 10 nm separation, the peaks of the spectrum experience a redshift and decrease significantly. For slabs coated with graphene sheets at 50 nm separation, the peaks of the spectrum also show a slight redshift but the high-frequency peak increases a little bit. When slabs are coated with graphene sheets, the whole spectrum (apart from the peaks) increases by about two orders of magnitude. Furthermore, when patterning graphene sheets into gratings, almost the whole spectrum (apart from the peaks) increases by further two orders of magnitude. We analyse now this mechanism. \\

\begin{figure*} [htbp]
\centerline {\includegraphics[width=0.8\textwidth]{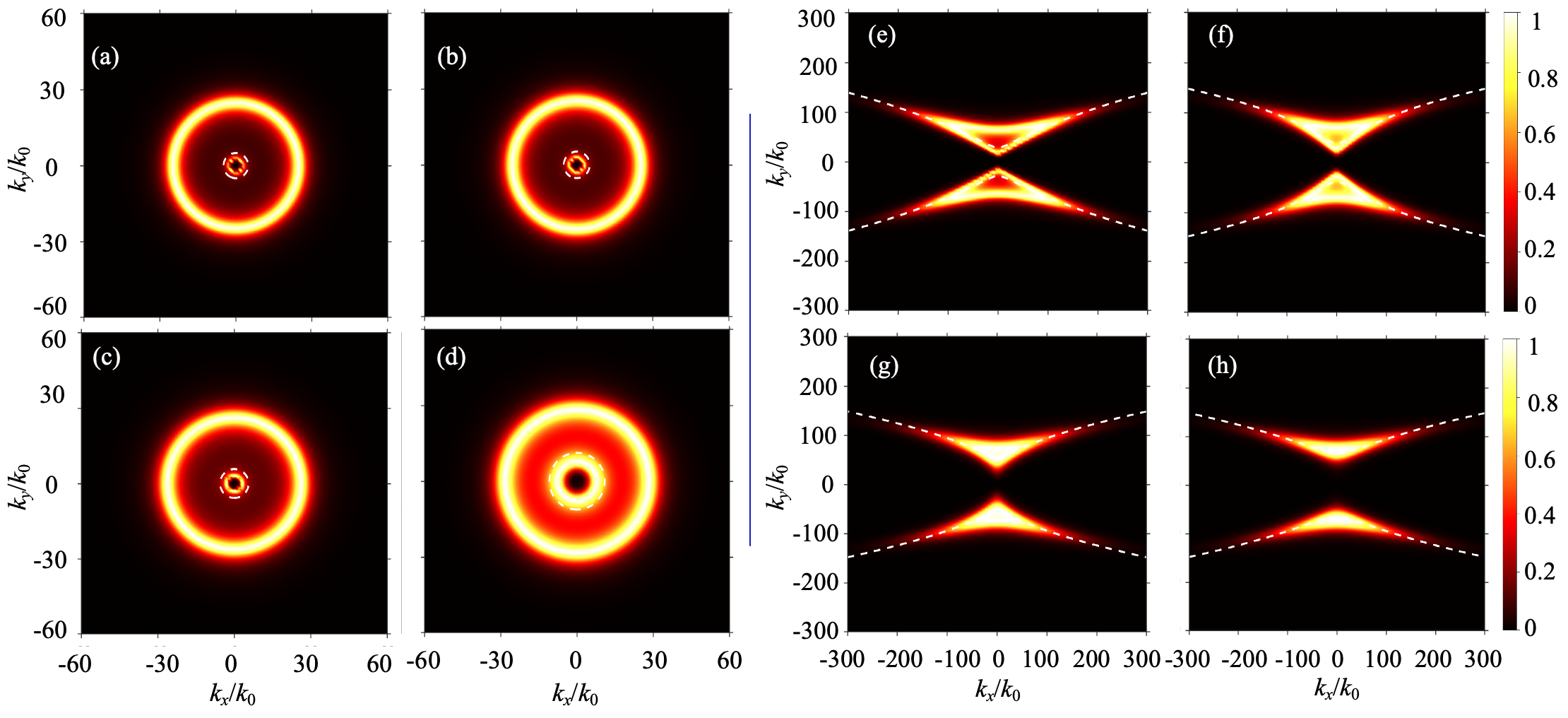}}
     \caption{Energy transmission coefficient for a fused silica substrate coated with a graphene sheet (and graphene grating) considering different substrate thicknesses: (a, e) $h=20$ nm; (b, f) $h=50$ nm; (c, g) $h=100$ nm; (d, h) $h=500$ nm. Panels (a)-(b)-(c)-(d) are for graphene sheet coating cases. Panels (e)-(f)-(g)- (h) are for the graphene grating coating cases. Here : $d=50$ nm, $\mu$=0.5 eV and $\omega=5\times10^{13}$ rad/s. The dotted curves represent the dispersion relations.}
   \label{fig:circular_mode}
\end{figure*}

In Fig.~\ref{fig:circular_mode}, we show the energy transmission coefficient (which is the sum over the two polarisations of the photon tunneling probabilities, at a fixed angular frequency $\omega$ and lateral wavevector $\textbf{k}$ \cite{Greffet2012prb_trans_def,Liu2014apl_trans_def}) in the ($k_x/k_0$,$k_y/k_0$) plane at the angular frequency $\omega=5\times10^{13}$ rad/s for a slab coated with a graphene sheet (Figs.~\ref{fig:circular_mode} (a) - (d)) and for a  slab coated with a graphene grating (Figs.~\ref{fig:circular_mode} (e) - (h)) respectively.  When patterning the graphene sheet into a grating, the plasmon experiences a topological transition from circular to hyperbolic one. The dotted curves represent the dispersion relations, that can be obtained from the poles of the reflection coefficient \cite{Zhou2022langmuir,Zhou2022prm}. For the graphene grating case, there are more accessible high-$k$ modes, while for the graphene sheet configuration, the accessible wavevector region is relatively smaller than that of the graphene grating one. Consequently, the slabs coated with a graphene grating can facilitate a greater transfer of energy compared to the slabs coated with a graphene sheet. 


\begin{figure} [htbp]
\centerline {\includegraphics[width=0.4\textwidth]{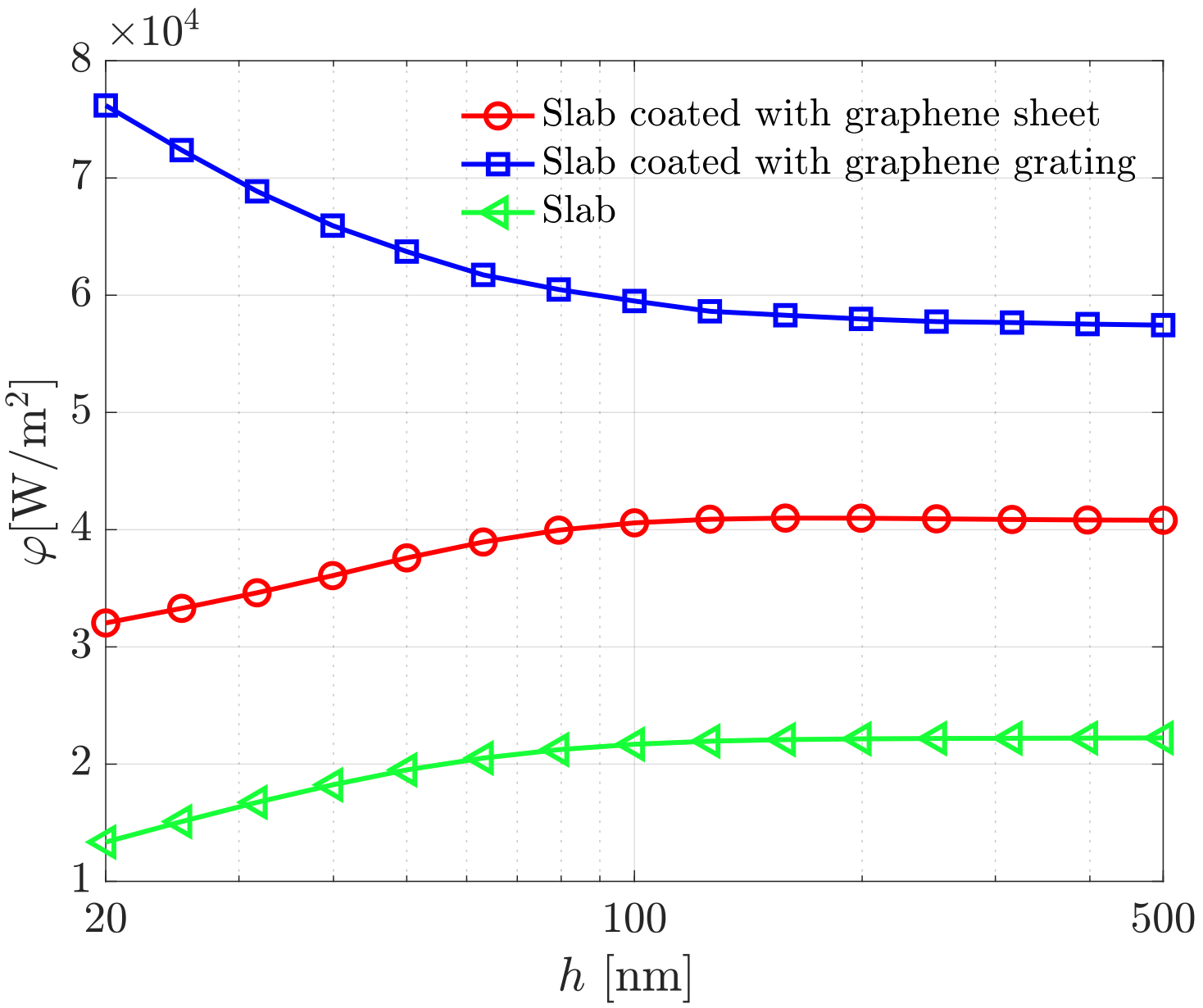}}
     \caption{The dependence of the radiative heat flux $\varphi$ on the thickness $h$. The separation distance is $d=50$ nm.}
   \label{fig:thickness_dependence}
\end{figure}
Furthermore, when we change the thickness of the substrate, in the range 20 nm to 500 nm, the accessible region for the high-value energy transmission coefficient increases a little bit for the slab coated with the graphene sheet [Fig.~\ref{fig:circular_mode} panels (a)-(b)-(c)-(d)], while it decreases slightly for the slab coated with the graphene grating [Fig.~\ref{fig:circular_mode} panels (e)-(f)-(g)-(h)]. Thus, we can predict that in the case of a slab coated with a graphene sheet, increasing the substrate thickness can enhance the energy transfer. However, for a slab coated with a graphene grating, the opposite effect occurs; increasing substrate thickness will hinder energy transfer. This is confirmed in Fig.~\ref{fig:thickness_dependence} where we show the numerical results for the dependence of the radiative heat flux $\varphi$ on the substrate thickness $h$. 

\begin{figure} [htbp]
\centerline {\includegraphics[width=0.4\textwidth]{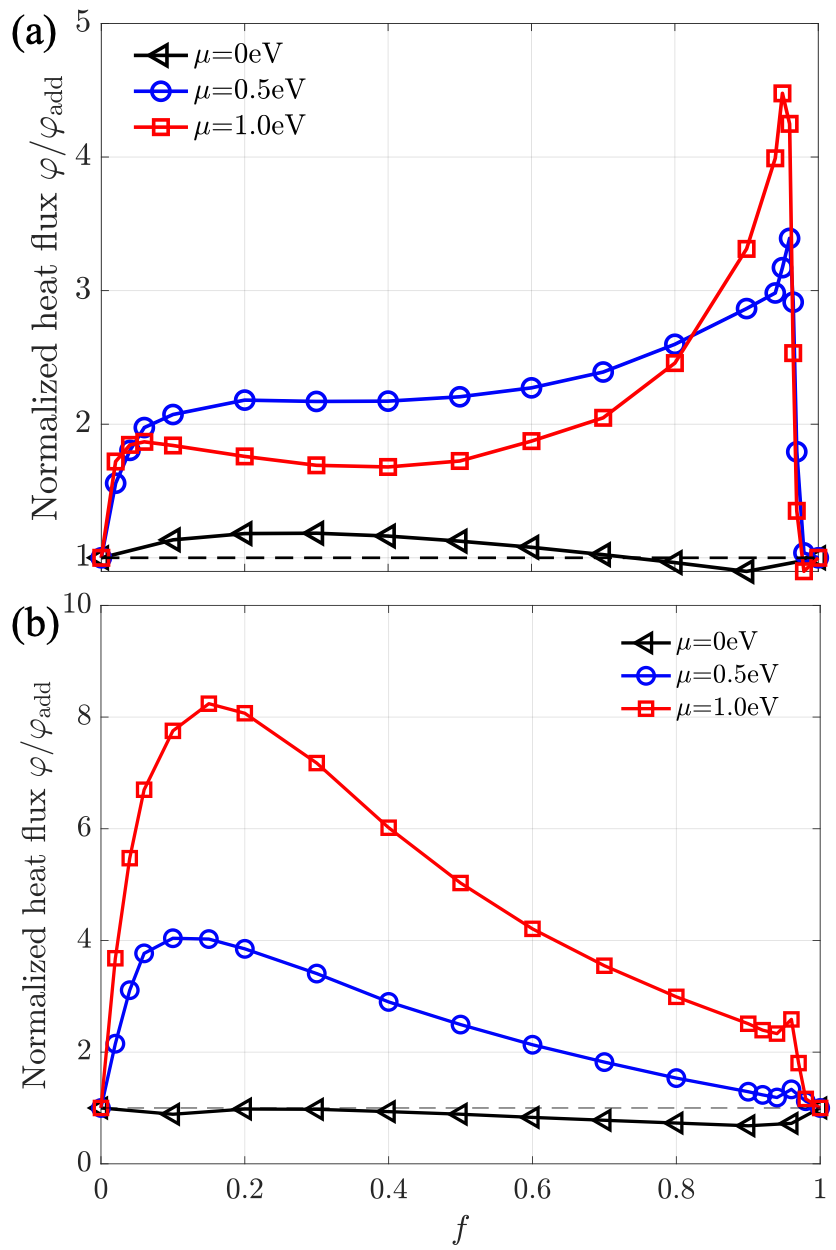}}
     \caption{The dependence of normalized heat flux $\varphi/\varphi_{\rm add}$ on the filling fraction $f$ for fused silica slabs coated with graphene grating considering two different separation distances: (a) $d=10$ nm; (b) $d=100$ nm. $\mu=0$ eV, 0.5 eV, and 1.0 eV. The grating period $D$ is 20 nm.}
   \label{fig:Nonadditivity}
\end{figure}

We investigate now a possible approximation of the near-field radiative heat flux $\varphi$ of Eq.~(\ref{HF}) between slabs coated with graphene gratings. To this extent we introduce an approximate additive expression $\varphi_{\rm add}$ of the flux given by a sum of the flux $\varphi_{\rm bs}$ occurring between bare substrates and the flux $\varphi_{\rm sh}$ occurring between slabs completely covered with a graphene sheets, weighted by the grating filling fraction $f$: 
\begin{equation}
\varphi_{\rm add}=(1-f)\varphi_{\rm bs}+f\varphi_{\rm sh},
\label{HF_add}
\end{equation}
 In Fig.~\ref{fig:Nonadditivity} we show the the normalized radiative heat flux $\varphi/\varphi_{\rm add}$ as a function of the filling fraction $f$, and for two different separation distances, (a) $d=10$ nm, and (b) $d=100$ nm, and different values of the chemical potentials $\mu=0$ eV, 0.5 eV, and 1.0 eV. 

First, Fig.~\ref{fig:Nonadditivity} clearly shows that for a fixed separation distance, the dependence of $\varphi/\varphi_{\rm add}$ on the filling fraction $f$ is strongly dependent on the chemical potential. Second, when $\mu$ = 0 eV, $\varphi/\varphi_{\rm add}$ is almost equal to 1 for $d=10$nm and close to it for $d=100$nm. This means that we are in a region of additivity of the radiative heat transfer and one can use the simple additive approximation given by Eq.~(\ref{HF_add}) safely. It is of fundamental importance to measure the advantage of this formula over exact calculations in terms of computational resources. For $\mu$ = 1 eV, the $\varphi/\varphi_{\rm add}$ reaches its maximum values of 4.5 and 8.2 for separation distances $d=10$ nm and 100 nm, respectively. The radiative heat transfer is then strongly non-additive and this non-additivity is more pronounced for $d=100$nm than that for $d=10$nm where the maximum mismatch nearly doubles. This is due to the fact that for larger distances the scattering details become more important (the nanostructures see each other in more details), where a full theory including the complexity of the nanostructures is needed in the comparison with the experiments. In addition, when the filling fraction $f$ approaches 0 or 1 and the slab coated with graphene grating transits to the bare slab or to the slab fully coated with a graphene sheet, the $\varphi/\varphi_{\rm add}$ drops dramatically to 1. When the filling fraction $f$ deviates from 0 or 1, the graphene grating coating facilitates the excitation of high-$k$ hyperbolic modes, enabling efficient energy transfer. Finally, we stress that this structure has also the remarkable feature that the near-field radiative heat transfer is here adjustable \textit{in situ} by merely changing the chemical potential (which can be made {\it e.g.} via a gate voltage applied to the structure). This capacity to modulate NFRHT without necessitating any geometric alterations holds substantial importance for experimental investigations.

In summary, the near-field radiative heat transfer between finite-thickness planar fused silica slabs coated with graphene gratings is analysed by using an exact Fourier modal method equipped with local basis functions (FMM-LBF). This approach is beyond the effective medium approximation, hence being relevant for realistic experimental analysis. In previous studies, it has been shown that  coating a substrate with a full graphene sheet generally decreases the NFRHT at short separations (typically for d<100 nm) {compared to the bare substrates}. In this work, we focused on these short separations, where the EMT cannot be used since it significantly overestimates the radiative heat flux (relative error to the exact one can be as large as 50$\%$). We show that by patterning the graphene sheet coating into a strips grating coating, {a topological transition occurs in the energy transmission coefficient diagrams of the considered graphene-nanostructure (changing from circular to hyperbolic),} allowing for more channels for energy transfer. Hence, the radiative energy flux between the substrates coated with graphene gratings is much higher than the one between slabs fully covered with graphene sheets, as well as the one between uncoated substrates. We compare the exact calculation with an approximate additive one. In general, the radiative heat flux between graphene-grating coated slabs deviates significantly from the additive results. However, for low chemical potentials, the additive approximation is quite good. A significant dependence of the radiative heat transfer on the chemical potential is found, which has potential to \textit{in situ} modulate the scattering features of the graphene grating without any geometric alterations. This work has the potential to unveil new avenues for harnessing non-additive heat transfer effects in graphene-based nano-devices.

\begin{acknowledgments}
This work described was supported by a grant "CAT" (No. A-HKUST604/20) from the ANR/RGC Joint Research Scheme sponsored by the French National Research Agency (ANR) and the Research Grants Council (RGC) of the Hong Kong Special Administrative Region, China.
\end{acknowledgments}

\bibliography{HT}

\providecommand{\noopsort}[1]{}\providecommand{\singleletter}[1]{#1}%
\begin{thebibliography}{63}%
\makeatletter
\providecommand \@ifxundefined [1]{%
 \@ifx{#1\undefined}
}%
\providecommand \@ifnum [1]{%
 \ifnum #1\expandafter \@firstoftwo
 \else \expandafter \@secondoftwo
 \fi
}%
\providecommand \@ifx [1]{%
 \ifx #1\expandafter \@firstoftwo
 \else \expandafter \@secondoftwo
 \fi
}%
\providecommand \natexlab [1]{#1}%
\providecommand \enquote  [1]{``#1''}%
\providecommand \bibnamefont  [1]{#1}%
\providecommand \bibfnamefont [1]{#1}%
\providecommand \citenamefont [1]{#1}%
\providecommand \href@noop [0]{\@secondoftwo}%
\providecommand \href [0]{\begingroup \@sanitize@url \@href}%
\providecommand \@href[1]{\@@startlink{#1}\@@href}%
\providecommand \@@href[1]{\endgroup#1\@@endlink}%
\providecommand \@sanitize@url [0]{\catcode `\\12\catcode `\$12\catcode
  `\&12\catcode `\#12\catcode `\^12\catcode `\_12\catcode `\%12\relax}%
\providecommand \@@startlink[1]{}%
\providecommand \@@endlink[0]{}%
\providecommand \url  [0]{\begingroup\@sanitize@url \@url }%
\providecommand \@url [1]{\endgroup\@href {#1}{\urlprefix }}%
\providecommand \urlprefix  [0]{URL }%
\providecommand \Eprint [0]{\href }%
\providecommand \doibase [0]{http://dx.doi.org/}%
\providecommand \selectlanguage [0]{\@gobble}%
\providecommand \bibinfo  [0]{\@secondoftwo}%
\providecommand \bibfield  [0]{\@secondoftwo}%
\providecommand \translation [1]{[#1]}%
\providecommand \BibitemOpen [0]{}%
\providecommand \bibitemStop [0]{}%
\providecommand \bibitemNoStop [0]{.\EOS\space}%
\providecommand \EOS [0]{\spacefactor3000\relax}%
\providecommand \BibitemShut  [1]{\csname bibitem#1\endcsname}%
\let\auto@bib@innerbib\@empty
\bibitem [{\citenamefont {Rytov}, \citenamefont {Kravtsov},\ and\ \citenamefont
  {Tatarskii}(1989)}]{Rytov1989}%
  \BibitemOpen
  \bibfield  {author} {\bibinfo {author} {\bibfnamefont {S.~M.}\ \bibnamefont
  {Rytov}}, \bibinfo {author} {\bibfnamefont {Y.~A.}\ \bibnamefont {Kravtsov}},
  \ and\ \bibinfo {author} {\bibfnamefont {V.~I.}\ \bibnamefont {Tatarskii}},\
  }\href@noop {} {\emph {\bibinfo {title} {Priniciples of statistical
  radiophysics}}},\ Vol.~\bibinfo {volume} {3}\ (\bibinfo  {publisher}
  {Springer-Verlag},\ \bibinfo {year} {1989})\BibitemShut {NoStop}%
\bibitem [{\citenamefont {Polder}\ and\ \citenamefont
  {Van~Hove}(1971)}]{Polder1971}%
  \BibitemOpen
  \bibfield  {author} {\bibinfo {author} {\bibfnamefont {D.}~\bibnamefont
  {Polder}}\ and\ \bibinfo {author} {\bibfnamefont {M.}~\bibnamefont
  {Van~Hove}},\ }\href {\doibase 10.1103/PhysRevB.4.3303} {\bibfield  {journal}
  {\bibinfo  {journal} {Phys. Rev. B}\ }\textbf {\bibinfo {volume} {4}},\
  \bibinfo {pages} {3303} (\bibinfo {year} {1971})}\BibitemShut {NoStop}%
\bibitem [{\citenamefont {Loomis}\ and\ \citenamefont
  {Maris}(1994)}]{Loomis1994}%
  \BibitemOpen
  \bibfield  {author} {\bibinfo {author} {\bibfnamefont {J.~J.}\ \bibnamefont
  {Loomis}}\ and\ \bibinfo {author} {\bibfnamefont {H.~J.}\ \bibnamefont
  {Maris}},\ }\href {\doibase 10.1103/PhysRevB.50.18517} {\bibfield  {journal}
  {\bibinfo  {journal} {Phys. Rev. B}\ }\textbf {\bibinfo {volume} {50}},\
  \bibinfo {pages} {18517} (\bibinfo {year} {1994})}\BibitemShut {NoStop}%
\bibitem [{\citenamefont {Carminati}\ and\ \citenamefont
  {Greffet}(1999)}]{Carminati1999}%
  \BibitemOpen
  \bibfield  {author} {\bibinfo {author} {\bibfnamefont {R.}~\bibnamefont
  {Carminati}}\ and\ \bibinfo {author} {\bibfnamefont {J.-J.}\ \bibnamefont
  {Greffet}},\ }\href {\doibase 10.1103/PhysRevLett.82.1660} {\bibfield
  {journal} {\bibinfo  {journal} {Phys. Rev. Lett.}\ }\textbf {\bibinfo
  {volume} {82}},\ \bibinfo {pages} {1660} (\bibinfo {year}
  {1999})}\BibitemShut {NoStop}%
\bibitem [{\citenamefont {Shchegrov}\ \emph {et~al.}(2000)\citenamefont
  {Shchegrov}, \citenamefont {Joulain}, \citenamefont {Carminati},\ and\
  \citenamefont {Greffet}}]{Shchegrov2000}%
  \BibitemOpen
  \bibfield  {author} {\bibinfo {author} {\bibfnamefont {A.~V.}\ \bibnamefont
  {Shchegrov}}, \bibinfo {author} {\bibfnamefont {K.}~\bibnamefont {Joulain}},
  \bibinfo {author} {\bibfnamefont {R.}~\bibnamefont {Carminati}}, \ and\
  \bibinfo {author} {\bibfnamefont {J.-J.}\ \bibnamefont {Greffet}},\ }\href
  {\doibase 10.1103/PhysRevLett.85.1548} {\bibfield  {journal} {\bibinfo
  {journal} {Phys. Rev. Lett.}\ }\textbf {\bibinfo {volume} {85}},\ \bibinfo
  {pages} {1548} (\bibinfo {year} {2000})}\BibitemShut {NoStop}%
\bibitem [{\citenamefont {Volokitin}\ and\ \citenamefont
  {Persson}(2001)}]{Volokitin2001}%
  \BibitemOpen
  \bibfield  {author} {\bibinfo {author} {\bibfnamefont {A.~I.}\ \bibnamefont
  {Volokitin}}\ and\ \bibinfo {author} {\bibfnamefont {B.~N.~J.}\ \bibnamefont
  {Persson}},\ }\href {\doibase 10.1103/PhysRevB.63.205404} {\bibfield
  {journal} {\bibinfo  {journal} {Phys. Rev. B}\ }\textbf {\bibinfo {volume}
  {63}},\ \bibinfo {pages} {205404} (\bibinfo {year} {2001})}\BibitemShut
  {NoStop}%
\bibitem [{\citenamefont {Narayanaswamy}\ and\ \citenamefont
  {Chen}(2008)}]{Narayanaswamy2008}%
  \BibitemOpen
  \bibfield  {author} {\bibinfo {author} {\bibfnamefont {A.}~\bibnamefont
  {Narayanaswamy}}\ and\ \bibinfo {author} {\bibfnamefont {G.}~\bibnamefont
  {Chen}},\ }\href {https://link.aps.org/doi/10.1103/PhysRevB.77.075125}
  {\bibfield  {journal} {\bibinfo  {journal} {Phys. Rev. B}\ }\textbf {\bibinfo
  {volume} {77}},\ \bibinfo {pages} {075125} (\bibinfo {year}
  {2008})}\BibitemShut {NoStop}%
\bibitem [{\citenamefont {Chapuis}\ \emph
  {et~al.}(2008{\natexlab{a}})\citenamefont {Chapuis}, \citenamefont {Laroche},
  \citenamefont {Volz},\ and\ \citenamefont {Greffet}}]{Chapuis2008}%
  \BibitemOpen
  \bibfield  {author} {\bibinfo {author} {\bibfnamefont {P.~O.}\ \bibnamefont
  {Chapuis}}, \bibinfo {author} {\bibfnamefont {M.}~\bibnamefont {Laroche}},
  \bibinfo {author} {\bibfnamefont {S.}~\bibnamefont {Volz}}, \ and\ \bibinfo
  {author} {\bibfnamefont {J.-J.}\ \bibnamefont {Greffet}},\ }\href {\doibase
  10.1063/1.2931062} {\bibfield  {journal} {\bibinfo  {journal} {Appl. Phys.
  Lett.}\ }\textbf {\bibinfo {volume} {92}},\ \bibinfo {pages} {3303} (\bibinfo
  {year} {2008}{\natexlab{a}})}\BibitemShut {NoStop}%
\bibitem [{\citenamefont {Manjavacas}\ and\ \citenamefont {Garc\'{\i}a~de
  Abajo}(2012)}]{Manjavacas2012}%
  \BibitemOpen
  \bibfield  {author} {\bibinfo {author} {\bibfnamefont {A.}~\bibnamefont
  {Manjavacas}}\ and\ \bibinfo {author} {\bibfnamefont {F.~J.}\ \bibnamefont
  {Garc\'{\i}a~de Abajo}},\ }\href {\doibase 10.1103/PhysRevB.86.075466}
  {\bibfield  {journal} {\bibinfo  {journal} {Phys. Rev. B}\ }\textbf {\bibinfo
  {volume} {86}},\ \bibinfo {pages} {075466} (\bibinfo {year}
  {2012})}\BibitemShut {NoStop}%
\bibitem [{\citenamefont {Nikbakht}(2018)}]{Nikbakht2018}%
  \BibitemOpen
  \bibfield  {author} {\bibinfo {author} {\bibfnamefont {M.}~\bibnamefont
  {Nikbakht}},\ }\href {\doibase https://doi.org/10.1016/j.jqsrt.2018.10.005}
  {\bibfield  {journal} {\bibinfo  {journal} {J. Quant. Spectrosc. Radiat.
  Transf.}\ }\textbf {\bibinfo {volume} {221}},\ \bibinfo {pages} {164}
  (\bibinfo {year} {2018})}\BibitemShut {NoStop}%
\bibitem [{\citenamefont {Messina}, \citenamefont {Biehs},\ and\ \citenamefont
  {Ben-Abdallah}(2018)}]{Messina2018}%
  \BibitemOpen
  \bibfield  {author} {\bibinfo {author} {\bibfnamefont {R.}~\bibnamefont
  {Messina}}, \bibinfo {author} {\bibfnamefont {S.-A.}\ \bibnamefont {Biehs}},
  \ and\ \bibinfo {author} {\bibfnamefont {P.}~\bibnamefont {Ben-Abdallah}},\
  }\href {\doibase 10.1103/PhysRevB.97.165437} {\bibfield  {journal} {\bibinfo
  {journal} {Phys. Rev. B}\ }\textbf {\bibinfo {volume} {97}},\ \bibinfo
  {pages} {165437} (\bibinfo {year} {2018})}\BibitemShut {NoStop}%
\bibitem [{\citenamefont {Dong}, \citenamefont {Zhao},\ and\ \citenamefont
  {Liu}(2018)}]{DongPrb2018}%
  \BibitemOpen
  \bibfield  {author} {\bibinfo {author} {\bibfnamefont {J.}~\bibnamefont
  {Dong}}, \bibinfo {author} {\bibfnamefont {J.~M.}\ \bibnamefont {Zhao}}, \
  and\ \bibinfo {author} {\bibfnamefont {L.~H.}\ \bibnamefont {Liu}},\ }\href
  {\doibase 10.1103/PhysRevB.97.075422} {\bibfield  {journal} {\bibinfo
  {journal} {Phys. Rev. B}\ }\textbf {\bibinfo {volume} {97}},\ \bibinfo
  {pages} {075422} (\bibinfo {year} {2018})}\BibitemShut {NoStop}%
\bibitem [{\citenamefont {Zhang}\ \emph
  {et~al.}(2019{\natexlab{a}})\citenamefont {Zhang}, \citenamefont {Antezza},
  \citenamefont {Yi},\ and\ \citenamefont {Tan}}]{Zhang2019R}%
  \BibitemOpen
  \bibfield  {author} {\bibinfo {author} {\bibfnamefont {Y.}~\bibnamefont
  {Zhang}}, \bibinfo {author} {\bibfnamefont {M.}~\bibnamefont {Antezza}},
  \bibinfo {author} {\bibfnamefont {H.~L.}\ \bibnamefont {Yi}}, \ and\ \bibinfo
  {author} {\bibfnamefont {H.~P.}\ \bibnamefont {Tan}},\ }\href {\doibase
  10.1103/PhysRevB.100.085426} {\bibfield  {journal} {\bibinfo  {journal}
  {Phys. Rev. B}\ }\textbf {\bibinfo {volume} {100}},\ \bibinfo {pages}
  {085426} (\bibinfo {year} {2019}{\natexlab{a}})}\BibitemShut {NoStop}%
\bibitem [{\citenamefont {He}\ \emph {et~al.}(2019)\citenamefont {He},
  \citenamefont {Qi}, \citenamefont {Ren}, \citenamefont {Zhao},\ and\
  \citenamefont {Antezza}}]{He2019APL}%
  \BibitemOpen
  \bibfield  {author} {\bibinfo {author} {\bibfnamefont {M.~J.}\ \bibnamefont
  {He}}, \bibinfo {author} {\bibfnamefont {H.}~\bibnamefont {Qi}}, \bibinfo
  {author} {\bibfnamefont {Y.~T.}\ \bibnamefont {Ren}}, \bibinfo {author}
  {\bibfnamefont {Y.~J.}\ \bibnamefont {Zhao}}, \ and\ \bibinfo {author}
  {\bibfnamefont {M.}~\bibnamefont {Antezza}},\ }\href {\doibase
  10.1063/1.5132995} {\bibfield  {journal} {\bibinfo  {journal} {Appl. Phys.
  Lett.}\ }\textbf {\bibinfo {volume} {115}},\ \bibinfo {pages} {263101}
  (\bibinfo {year} {2019})}\BibitemShut {NoStop}%
\bibitem [{\citenamefont {Zhang}\ \emph
  {et~al.}(2019{\natexlab{b}})\citenamefont {Zhang}, \citenamefont {Yi},
  \citenamefont {Tan},\ and\ \citenamefont {Antezza}}]{Zhang2019T}%
  \BibitemOpen
  \bibfield  {author} {\bibinfo {author} {\bibfnamefont {Y.}~\bibnamefont
  {Zhang}}, \bibinfo {author} {\bibfnamefont {H.~L.}\ \bibnamefont {Yi}},
  \bibinfo {author} {\bibfnamefont {H.~P.}\ \bibnamefont {Tan}}, \ and\
  \bibinfo {author} {\bibfnamefont {M.}~\bibnamefont {Antezza}},\ }\href
  {\doibase 10.1103/PhysRevB.100.134305} {\bibfield  {journal} {\bibinfo
  {journal} {Phys. Rev. B}\ }\textbf {\bibinfo {volume} {100}},\ \bibinfo
  {pages} {134305} (\bibinfo {year} {2019}{\natexlab{b}})}\BibitemShut
  {NoStop}%
\bibitem [{\citenamefont {Chapuis}\ \emph
  {et~al.}(2008{\natexlab{b}})\citenamefont {Chapuis}, \citenamefont {Laroche},
  \citenamefont {Volz},\ and\ \citenamefont {Greffet}}]{Chapuis2008plate}%
  \BibitemOpen
  \bibfield  {author} {\bibinfo {author} {\bibfnamefont {P.~O.}\ \bibnamefont
  {Chapuis}}, \bibinfo {author} {\bibfnamefont {M.}~\bibnamefont {Laroche}},
  \bibinfo {author} {\bibfnamefont {S.}~\bibnamefont {Volz}}, \ and\ \bibinfo
  {author} {\bibfnamefont {J.-J.}\ \bibnamefont {Greffet}},\ }\href {\doibase
  10.1103/PhysRevB.77.125402} {\bibfield  {journal} {\bibinfo  {journal} {Phys.
  Rev. B}\ }\textbf {\bibinfo {volume} {77}},\ \bibinfo {pages} {125402}
  (\bibinfo {year} {2008}{\natexlab{b}})}\BibitemShut {NoStop}%
\bibitem [{\citenamefont {Yang}\ and\ \citenamefont
  {Wang}(2016)}]{Yang2017prl}%
  \BibitemOpen
  \bibfield  {author} {\bibinfo {author} {\bibfnamefont {Y.}~\bibnamefont
  {Yang}}\ and\ \bibinfo {author} {\bibfnamefont {L.}~\bibnamefont {Wang}},\
  }\href {\doibase 10.1103/PhysRevLett.117.044301} {\bibfield  {journal}
  {\bibinfo  {journal} {Phys. Rev. Lett.}\ }\textbf {\bibinfo {volume} {117}},\
  \bibinfo {pages} {044301} (\bibinfo {year} {2016})}\BibitemShut {NoStop}%
\bibitem [{\citenamefont {Liu}\ \emph {et~al.}(2022{\natexlab{a}})\citenamefont
  {Liu}, \citenamefont {Chen}, \citenamefont {Caratenuto}, \citenamefont
  {Tian}, \citenamefont {Liu}, \citenamefont {Zhao},\ and\ \citenamefont
  {Zheng}}]{Zheng2022Materials_EMA}%
  \BibitemOpen
  \bibfield  {author} {\bibinfo {author} {\bibfnamefont {Y.}~\bibnamefont
  {Liu}}, \bibinfo {author} {\bibfnamefont {F.~Q.}\ \bibnamefont {Chen}},
  \bibinfo {author} {\bibfnamefont {A.}~\bibnamefont {Caratenuto}}, \bibinfo
  {author} {\bibfnamefont {Y.~P.}\ \bibnamefont {Tian}}, \bibinfo {author}
  {\bibfnamefont {X.~J.}\ \bibnamefont {Liu}}, \bibinfo {author} {\bibfnamefont
  {Y.~T.}\ \bibnamefont {Zhao}}, \ and\ \bibinfo {author} {\bibfnamefont
  {Y.}~\bibnamefont {Zheng}},\ }\href {\doibase 10.3390/ma15030998} {\bibfield
  {journal} {\bibinfo  {journal} {Materials}\ }\textbf {\bibinfo {volume} {15}}
  (\bibinfo {year} {2022}{\natexlab{a}}),\ 10.3390/ma15030998}\BibitemShut
  {NoStop}%
\bibitem [{\citenamefont {Biehs}, \citenamefont {Rosa},\ and\ \citenamefont
  {Ben-Abdallah}(2011)}]{Biehs2011gratings}%
  \BibitemOpen
  \bibfield  {author} {\bibinfo {author} {\bibfnamefont {S.-A.}\ \bibnamefont
  {Biehs}}, \bibinfo {author} {\bibfnamefont {F.~S.~S.}\ \bibnamefont {Rosa}},
  \ and\ \bibinfo {author} {\bibfnamefont {P.}~\bibnamefont {Ben-Abdallah}},\
  }\href {\doibase 10.1063/1.3596707} {\bibfield  {journal} {\bibinfo
  {journal} {Appl. Phys. Lett.}\ }\textbf {\bibinfo {volume} {98}},\ \bibinfo
  {pages} {243102} (\bibinfo {year} {2011})}\BibitemShut {NoStop}%
\bibitem [{\citenamefont {Kan}, \citenamefont {Zhao},\ and\ \citenamefont
  {Zhang}(2019)}]{Kan2019prb}%
  \BibitemOpen
  \bibfield  {author} {\bibinfo {author} {\bibfnamefont {Y.~H.}\ \bibnamefont
  {Kan}}, \bibinfo {author} {\bibfnamefont {C.~Y.}\ \bibnamefont {Zhao}}, \
  and\ \bibinfo {author} {\bibfnamefont {Z.~M.}\ \bibnamefont {Zhang}},\ }\href
  {\doibase 10.1103/PhysRevB.99.035433} {\bibfield  {journal} {\bibinfo
  {journal} {Phys. Rev. B}\ }\textbf {\bibinfo {volume} {99}},\ \bibinfo
  {pages} {035433} (\bibinfo {year} {2019})}\BibitemShut {NoStop}%
\bibitem [{\citenamefont {Kan}\ and\ \citenamefont
  {Zhao}(2021)}]{Kan2021ijhmt}%
  \BibitemOpen
  \bibfield  {author} {\bibinfo {author} {\bibfnamefont {Y.}~\bibnamefont
  {Kan}}\ and\ \bibinfo {author} {\bibfnamefont {C.}~\bibnamefont {Zhao}},\
  }\href {\doibase https://doi.org/10.1016/j.ijheatmasstransfer.2021.121124}
  {\bibfield  {journal} {\bibinfo  {journal} {Int. J. Heat Mass Transf.}\
  }\textbf {\bibinfo {volume} {171}},\ \bibinfo {pages} {121124} (\bibinfo
  {year} {2021})}\BibitemShut {NoStop}%
\bibitem [{\citenamefont {Yang}, \citenamefont {Sabbaghi},\ and\ \citenamefont
  {Wang}(2017)}]{Yang2017ijhmt}%
  \BibitemOpen
  \bibfield  {author} {\bibinfo {author} {\bibfnamefont {Y.}~\bibnamefont
  {Yang}}, \bibinfo {author} {\bibfnamefont {P.}~\bibnamefont {Sabbaghi}}, \
  and\ \bibinfo {author} {\bibfnamefont {L.}~\bibnamefont {Wang}},\ }\href
  {\doibase https://doi.org/10.1016/j.ijheatmasstransfer.2016.12.061}
  {\bibfield  {journal} {\bibinfo  {journal} {Int. J. Heat Mass Transf.}\
  }\textbf {\bibinfo {volume} {108}},\ \bibinfo {pages} {851} (\bibinfo {year}
  {2017})}\BibitemShut {NoStop}%
\bibitem [{\citenamefont {Ottens}\ \emph {et~al.}(2011)\citenamefont {Ottens},
  \citenamefont {Quetschke}, \citenamefont {Wise}, \citenamefont {Alemi},
  \citenamefont {Lundock}, \citenamefont {Mueller}, \citenamefont {Reitze},
  \citenamefont {Tanner},\ and\ \citenamefont {Whiting}}]{Ottens2011}%
  \BibitemOpen
  \bibfield  {author} {\bibinfo {author} {\bibfnamefont {R.~S.}\ \bibnamefont
  {Ottens}}, \bibinfo {author} {\bibfnamefont {V.}~\bibnamefont {Quetschke}},
  \bibinfo {author} {\bibfnamefont {S.}~\bibnamefont {Wise}}, \bibinfo {author}
  {\bibfnamefont {A.~A.}\ \bibnamefont {Alemi}}, \bibinfo {author}
  {\bibfnamefont {R.}~\bibnamefont {Lundock}}, \bibinfo {author} {\bibfnamefont
  {G.}~\bibnamefont {Mueller}}, \bibinfo {author} {\bibfnamefont {D.~H.}\
  \bibnamefont {Reitze}}, \bibinfo {author} {\bibfnamefont {D.~B.}\
  \bibnamefont {Tanner}}, \ and\ \bibinfo {author} {\bibfnamefont {B.~F.}\
  \bibnamefont {Whiting}},\ }\href {\doibase 10.1103/PhysRevLett.107.014301}
  {\bibfield  {journal} {\bibinfo  {journal} {Phys. Rev. Lett.}\ }\textbf
  {\bibinfo {volume} {107}},\ \bibinfo {pages} {014301} (\bibinfo {year}
  {2011})}\BibitemShut {NoStop}%
\bibitem [{\citenamefont {Lim}, \citenamefont {Lee},\ and\ \citenamefont
  {Lee}(2015)}]{Lim2015}%
  \BibitemOpen
  \bibfield  {author} {\bibinfo {author} {\bibfnamefont {M.}~\bibnamefont
  {Lim}}, \bibinfo {author} {\bibfnamefont {S.~S.}\ \bibnamefont {Lee}}, \ and\
  \bibinfo {author} {\bibfnamefont {B.~J.}\ \bibnamefont {Lee}},\ }\href
  {\doibase 10.1103/PhysRevB.91.195136} {\bibfield  {journal} {\bibinfo
  {journal} {Phys. Rev. B}\ }\textbf {\bibinfo {volume} {91}},\ \bibinfo
  {pages} {195136} (\bibinfo {year} {2015})}\BibitemShut {NoStop}%
\bibitem [{\citenamefont {Watjen}, \citenamefont {Zhao},\ and\ \citenamefont
  {Zhang}(2016)}]{Watjen2016}%
  \BibitemOpen
  \bibfield  {author} {\bibinfo {author} {\bibfnamefont {J.~I.}\ \bibnamefont
  {Watjen}}, \bibinfo {author} {\bibfnamefont {B.}~\bibnamefont {Zhao}}, \ and\
  \bibinfo {author} {\bibfnamefont {Z.~M.}\ \bibnamefont {Zhang}},\ }\href
  {https://doi.org/10.1063/1.4967384} {\bibfield  {journal} {\bibinfo
  {journal} {Appl. Phys. Lett.}\ }\textbf {\bibinfo {volume} {109}} (\bibinfo
  {year} {2016})}\BibitemShut {NoStop}%
\bibitem [{\citenamefont {Ghashami}\ \emph {et~al.}(2018)\citenamefont
  {Ghashami}, \citenamefont {Geng}, \citenamefont {Kim}, \citenamefont
  {Iacopino}, \citenamefont {Cho},\ and\ \citenamefont {Park}}]{Ghashami2018}%
  \BibitemOpen
  \bibfield  {author} {\bibinfo {author} {\bibfnamefont {M.}~\bibnamefont
  {Ghashami}}, \bibinfo {author} {\bibfnamefont {H.}~\bibnamefont {Geng}},
  \bibinfo {author} {\bibfnamefont {T.}~\bibnamefont {Kim}}, \bibinfo {author}
  {\bibfnamefont {N.}~\bibnamefont {Iacopino}}, \bibinfo {author}
  {\bibfnamefont {S.~K.}\ \bibnamefont {Cho}}, \ and\ \bibinfo {author}
  {\bibfnamefont {K.}~\bibnamefont {Park}},\ }\href {\doibase
  10.1103/PhysRevLett.120.175901} {\bibfield  {journal} {\bibinfo  {journal}
  {Phys. Rev. Lett.}\ }\textbf {\bibinfo {volume} {120}},\ \bibinfo {pages}
  {175901} (\bibinfo {year} {2018})}\BibitemShut {NoStop}%
\bibitem [{\citenamefont {Yang}\ \emph {et~al.}(2018)\citenamefont {Yang},
  \citenamefont {Du}, \citenamefont {Su}, \citenamefont {Fu}, \citenamefont
  {Gong}, \citenamefont {He},\ and\ \citenamefont {Ma}}]{Yungui2018NC}%
  \BibitemOpen
  \bibfield  {author} {\bibinfo {author} {\bibfnamefont {J.}~\bibnamefont
  {Yang}}, \bibinfo {author} {\bibfnamefont {W.}~\bibnamefont {Du}}, \bibinfo
  {author} {\bibfnamefont {Y.~S.}\ \bibnamefont {Su}}, \bibinfo {author}
  {\bibfnamefont {Y.}~\bibnamefont {Fu}}, \bibinfo {author} {\bibfnamefont
  {S.~X.}\ \bibnamefont {Gong}}, \bibinfo {author} {\bibfnamefont {S.~L.}\
  \bibnamefont {He}}, \ and\ \bibinfo {author} {\bibfnamefont {Y.~G.}\
  \bibnamefont {Ma}},\ }\href {\doibase 10.1038/s41467-018-06163-8} {\bibfield
  {journal} {\bibinfo  {journal} {Nat. Commun.}\ }\textbf {\bibinfo {volume}
  {9}},\ \bibinfo {pages} {4033} (\bibinfo {year} {2018})}\BibitemShut
  {NoStop}%
\bibitem [{\citenamefont {Iqbal}\ \emph {et~al.}(2023)\citenamefont {Iqbal},
  \citenamefont {Zhang}, \citenamefont {Wang}, \citenamefont {Fang},
  \citenamefont {Hu}, \citenamefont {Dang}, \citenamefont {Zhang},
  \citenamefont {Jin}, \citenamefont {Xu}, \citenamefont {Ju},\ and\
  \citenamefont {Ma}}]{Yungui2023PRAppl}%
  \BibitemOpen
  \bibfield  {author} {\bibinfo {author} {\bibfnamefont {N.}~\bibnamefont
  {Iqbal}}, \bibinfo {author} {\bibfnamefont {S.}~\bibnamefont {Zhang}},
  \bibinfo {author} {\bibfnamefont {S.}~\bibnamefont {Wang}}, \bibinfo {author}
  {\bibfnamefont {Z.~Z.}\ \bibnamefont {Fang}}, \bibinfo {author}
  {\bibfnamefont {Y.~Y.}\ \bibnamefont {Hu}}, \bibinfo {author} {\bibfnamefont
  {Y.~D.}\ \bibnamefont {Dang}}, \bibinfo {author} {\bibfnamefont {M.~J.}\
  \bibnamefont {Zhang}}, \bibinfo {author} {\bibfnamefont {Y.}~\bibnamefont
  {Jin}}, \bibinfo {author} {\bibfnamefont {J.~B.}\ \bibnamefont {Xu}},
  \bibinfo {author} {\bibfnamefont {B.~F.}\ \bibnamefont {Ju}}, \ and\ \bibinfo
  {author} {\bibfnamefont {Y.~G.}\ \bibnamefont {Ma}},\ }\href {\doibase
  10.1103/PhysRevApplied.19.024019} {\bibfield  {journal} {\bibinfo  {journal}
  {Phys. Rev. Appl.}\ }\textbf {\bibinfo {volume} {19}},\ \bibinfo {pages}
  {024019} (\bibinfo {year} {2023})}\BibitemShut {NoStop}%
\bibitem [{\citenamefont {DeSutter}, \citenamefont {Tang},\ and\ \citenamefont
  {Francoeur}(2019)}]{DeSutter2019}%
  \BibitemOpen
  \bibfield  {author} {\bibinfo {author} {\bibfnamefont {J.}~\bibnamefont
  {DeSutter}}, \bibinfo {author} {\bibfnamefont {L.}~\bibnamefont {Tang}}, \
  and\ \bibinfo {author} {\bibfnamefont {M.}~\bibnamefont {Francoeur}},\ }\href
  {\doibase 10.1038/s41565-019-0483-1} {\bibfield  {journal} {\bibinfo
  {journal} {Nat. Nanotechnol.}\ }\textbf {\bibinfo {volume} {14}},\ \bibinfo
  {pages} {751} (\bibinfo {year} {2019})}\BibitemShut {NoStop}%
\bibitem [{\citenamefont {Shen}, \citenamefont {Narayanaswamy},\ and\
  \citenamefont {Chen}(2009)}]{Shen2009}%
  \BibitemOpen
  \bibfield  {author} {\bibinfo {author} {\bibfnamefont {S.}~\bibnamefont
  {Shen}}, \bibinfo {author} {\bibfnamefont {A.}~\bibnamefont {Narayanaswamy}},
  \ and\ \bibinfo {author} {\bibfnamefont {G.}~\bibnamefont {Chen}},\ }\href
  {\doibase 10.1021/nl901208v} {\bibfield  {journal} {\bibinfo  {journal} {Nano
  Lett.}\ }\textbf {\bibinfo {volume} {9}},\ \bibinfo {pages} {2909} (\bibinfo
  {year} {2009})}\BibitemShut {NoStop}%
\bibitem [{\citenamefont {Rousseau}\ \emph {et~al.}(2009)\citenamefont
  {Rousseau}, \citenamefont {Siria}, \citenamefont {Jourdan}, \citenamefont
  {Volz}, \citenamefont {Comin}, \citenamefont {Chevrier},\ and\ \citenamefont
  {Greffet}}]{Rousseau2009}%
  \BibitemOpen
  \bibfield  {author} {\bibinfo {author} {\bibfnamefont {E.}~\bibnamefont
  {Rousseau}}, \bibinfo {author} {\bibfnamefont {A.}~\bibnamefont {Siria}},
  \bibinfo {author} {\bibfnamefont {G.}~\bibnamefont {Jourdan}}, \bibinfo
  {author} {\bibfnamefont {S.}~\bibnamefont {Volz}}, \bibinfo {author}
  {\bibfnamefont {F.}~\bibnamefont {Comin}}, \bibinfo {author} {\bibfnamefont
  {J.}~\bibnamefont {Chevrier}}, \ and\ \bibinfo {author} {\bibfnamefont
  {J.-J.}\ \bibnamefont {Greffet}},\ }\href {\doibase
  http://www.nature.com/nphoton/journal/v3/n9/suppinfo/nphoton.2009.144_S1.html}
  {\bibfield  {journal} {\bibinfo  {journal} {Nat. Photonics}\ }\textbf
  {\bibinfo {volume} {3}},\ \bibinfo {pages} {514} (\bibinfo {year}
  {2009})}\BibitemShut {NoStop}%
\bibitem [{\citenamefont {Song}\ \emph {et~al.}(2015)\citenamefont {Song},
  \citenamefont {Ganjeh}, \citenamefont {Sadat}, \citenamefont {Thompson},
  \citenamefont {Fiorino}, \citenamefont {Fernández-Hurtado}, \citenamefont
  {Feist}, \citenamefont {Garcia-Vidal}, \citenamefont {Cuevas}, \citenamefont
  {Reddy},\ and\ \citenamefont {Meyhofer}}]{Song2015}%
  \BibitemOpen
  \bibfield  {author} {\bibinfo {author} {\bibfnamefont {B.}~\bibnamefont
  {Song}}, \bibinfo {author} {\bibfnamefont {Y.}~\bibnamefont {Ganjeh}},
  \bibinfo {author} {\bibfnamefont {S.}~\bibnamefont {Sadat}}, \bibinfo
  {author} {\bibfnamefont {D.}~\bibnamefont {Thompson}}, \bibinfo {author}
  {\bibfnamefont {A.}~\bibnamefont {Fiorino}}, \bibinfo {author} {\bibfnamefont
  {V.}~\bibnamefont {Fernández-Hurtado}}, \bibinfo {author} {\bibfnamefont
  {J.}~\bibnamefont {Feist}}, \bibinfo {author} {\bibfnamefont {F.~J.}\
  \bibnamefont {Garcia-Vidal}}, \bibinfo {author} {\bibfnamefont {J.~C.}\
  \bibnamefont {Cuevas}}, \bibinfo {author} {\bibfnamefont {P.}~\bibnamefont
  {Reddy}}, \ and\ \bibinfo {author} {\bibfnamefont {E.}~\bibnamefont
  {Meyhofer}},\ }\href {\doibase 10.1038/nnano.2015.6
  https://www.nature.com/articles/nnano.2015.6#supplementary-information}
  {\bibfield  {journal} {\bibinfo  {journal} {Nat. Nanotechnol.}\ }\textbf
  {\bibinfo {volume} {10}},\ \bibinfo {pages} {253} (\bibinfo {year}
  {2015})}\BibitemShut {NoStop}%
\bibitem [{\citenamefont {Lu}\ \emph {et~al.}(2022)\citenamefont {Lu},
  \citenamefont {Zhang}, \citenamefont {Ou}, \citenamefont {Li}, \citenamefont
  {Zhou}, \citenamefont {Song}, \citenamefont {Luo},\ and\ \citenamefont
  {Cheng}}]{Lu2022small}%
  \BibitemOpen
  \bibfield  {author} {\bibinfo {author} {\bibfnamefont {L.}~\bibnamefont
  {Lu}}, \bibinfo {author} {\bibfnamefont {B.}~\bibnamefont {Zhang}}, \bibinfo
  {author} {\bibfnamefont {H.}~\bibnamefont {Ou}}, \bibinfo {author}
  {\bibfnamefont {B.~W.}\ \bibnamefont {Li}}, \bibinfo {author} {\bibfnamefont
  {K.}~\bibnamefont {Zhou}}, \bibinfo {author} {\bibfnamefont {J.~L.}\
  \bibnamefont {Song}}, \bibinfo {author} {\bibfnamefont {Z.~X.}\ \bibnamefont
  {Luo}}, \ and\ \bibinfo {author} {\bibfnamefont {Q.}~\bibnamefont {Cheng}},\
  }\href {\doibase https://doi.org/10.1002/smll.202108032} {\bibfield
  {journal} {\bibinfo  {journal} {Small}\ }\textbf {\bibinfo {volume} {18}},\
  \bibinfo {pages} {2108032} (\bibinfo {year} {2022})}\BibitemShut {NoStop}%
\bibitem [{\citenamefont {Shi}\ \emph {et~al.}(2021)\citenamefont {Shi},
  \citenamefont {Chen}, \citenamefont {Xu}, \citenamefont {Evans},\ and\
  \citenamefont {He}}]{Shi2021am}%
  \BibitemOpen
  \bibfield  {author} {\bibinfo {author} {\bibfnamefont {K.~Z.}\ \bibnamefont
  {Shi}}, \bibinfo {author} {\bibfnamefont {Z.~Y.}\ \bibnamefont {Chen}},
  \bibinfo {author} {\bibfnamefont {X.}~\bibnamefont {Xu}}, \bibinfo {author}
  {\bibfnamefont {J.}~\bibnamefont {Evans}}, \ and\ \bibinfo {author}
  {\bibfnamefont {S.~L.}\ \bibnamefont {He}},\ }\href {\doibase
  https://doi.org/10.1002/adma.202106097} {\bibfield  {journal} {\bibinfo
  {journal} {Advanced Materials}\ }\textbf {\bibinfo {volume} {33}},\ \bibinfo
  {pages} {2106097} (\bibinfo {year} {2021})}\BibitemShut {NoStop}%
\bibitem [{\citenamefont {Volokitin}(2017)}]{Volokitin2017Dey}%
  \BibitemOpen
  \bibfield  {author} {\bibinfo {author} {\bibfnamefont {A.}~\bibnamefont
  {Volokitin}},\ }\href {\doibase doi:10.1515/zna-2016-0367} {\bibfield
  {journal} {\bibinfo  {journal} {Z. Naturforsch. A}\ }\textbf {\bibinfo
  {volume} {72}},\ \bibinfo {pages} {171} (\bibinfo {year} {2017})}\BibitemShut
  {NoStop}%
\bibitem [{\citenamefont {Zhao}\ \emph {et~al.}(2017)\citenamefont {Zhao},
  \citenamefont {Guizal}, \citenamefont {Zhang}, \citenamefont {Fan},\ and\
  \citenamefont {Antezza}}]{Zhao2017prb}%
  \BibitemOpen
  \bibfield  {author} {\bibinfo {author} {\bibfnamefont {B.}~\bibnamefont
  {Zhao}}, \bibinfo {author} {\bibfnamefont {B.}~\bibnamefont {Guizal}},
  \bibinfo {author} {\bibfnamefont {Z.~M.}\ \bibnamefont {Zhang}}, \bibinfo
  {author} {\bibfnamefont {S.}~\bibnamefont {Fan}}, \ and\ \bibinfo {author}
  {\bibfnamefont {M.}~\bibnamefont {Antezza}},\ }\href {\doibase
  10.1103/PhysRevB.95.245437} {\bibfield  {journal} {\bibinfo  {journal} {Phys.
  Rev. B}\ }\textbf {\bibinfo {volume} {95}},\ \bibinfo {pages} {245437}
  (\bibinfo {year} {2017})}\BibitemShut {NoStop}%
\bibitem [{\citenamefont {He}\ \emph {et~al.}(2020{\natexlab{a}})\citenamefont
  {He}, \citenamefont {Qi}, \citenamefont {Ren}, \citenamefont {Zhao},\ and\
  \citenamefont {Antezza}}]{He2020ijhmt}%
  \BibitemOpen
  \bibfield  {author} {\bibinfo {author} {\bibfnamefont {M.~J.}\ \bibnamefont
  {He}}, \bibinfo {author} {\bibfnamefont {H.}~\bibnamefont {Qi}}, \bibinfo
  {author} {\bibfnamefont {Y.-T.}\ \bibnamefont {Ren}}, \bibinfo {author}
  {\bibfnamefont {Y.-J.}\ \bibnamefont {Zhao}}, \ and\ \bibinfo {author}
  {\bibfnamefont {M.}~\bibnamefont {Antezza}},\ }\href {\doibase
  https://doi.org/10.1016/j.ijheatmasstransfer.2020.119305} {\bibfield
  {journal} {\bibinfo  {journal} {Int. J. Heat Mass Transf.}\ }\textbf
  {\bibinfo {volume} {150}},\ \bibinfo {pages} {119305} (\bibinfo {year}
  {2020}{\natexlab{a}})}\BibitemShut {NoStop}%
\bibitem [{\citenamefont {He}\ \emph {et~al.}(2020{\natexlab{b}})\citenamefont
  {He}, \citenamefont {Qi}, \citenamefont {Ren}, \citenamefont {Zhao},\ and\
  \citenamefont {Antezza}}]{He2020ol}%
  \BibitemOpen
  \bibfield  {author} {\bibinfo {author} {\bibfnamefont {M.~J.}\ \bibnamefont
  {He}}, \bibinfo {author} {\bibfnamefont {H.}~\bibnamefont {Qi}}, \bibinfo
  {author} {\bibfnamefont {Y.~T.}\ \bibnamefont {Ren}}, \bibinfo {author}
  {\bibfnamefont {Y.}~\bibnamefont {Zhao}}, \ and\ \bibinfo {author}
  {\bibfnamefont {M.}~\bibnamefont {Antezza}},\ }\href {\doibase
  10.1364/OL.392371} {\bibfield  {journal} {\bibinfo  {journal} {Opt. Lett.}\
  }\textbf {\bibinfo {volume} {45}},\ \bibinfo {pages} {2914} (\bibinfo {year}
  {2020}{\natexlab{b}})}\BibitemShut {NoStop}%
\bibitem [{\citenamefont {Zheng}\ \emph {et~al.}(2017)\citenamefont {Zheng},
  \citenamefont {Liu}, \citenamefont {Wang},\ and\ \citenamefont
  {Xuan}}]{Zheng2017}%
  \BibitemOpen
  \bibfield  {author} {\bibinfo {author} {\bibfnamefont {Z.~H.}\ \bibnamefont
  {Zheng}}, \bibinfo {author} {\bibfnamefont {X.~L.}\ \bibnamefont {Liu}},
  \bibinfo {author} {\bibfnamefont {A.}~\bibnamefont {Wang}}, \ and\ \bibinfo
  {author} {\bibfnamefont {Y.~M.}\ \bibnamefont {Xuan}},\ }\href {\doibase
  https://doi.org/10.1016/j.ijheatmasstransfer.2017.01.107} {\bibfield
  {journal} {\bibinfo  {journal} {Int. J. Heat Mass Transf.}\ }\textbf
  {\bibinfo {volume} {109}},\ \bibinfo {pages} {63 } (\bibinfo {year}
  {2017})}\BibitemShut {NoStop}%
\bibitem [{\citenamefont {Ilic}\ \emph {et~al.}(2012)\citenamefont {Ilic},
  \citenamefont {Jablan}, \citenamefont {Joannopoulos}, \citenamefont
  {Celanovic}, \citenamefont {Buljan},\ and\ \citenamefont {Solja\ifmmode
  \check{c}\else \v{c}\fi{}i\ifmmode~\acute{c}\else
  \'{c}\fi{}}}]{Ognjen2012prb}%
  \BibitemOpen
  \bibfield  {author} {\bibinfo {author} {\bibfnamefont {O.}~\bibnamefont
  {Ilic}}, \bibinfo {author} {\bibfnamefont {M.}~\bibnamefont {Jablan}},
  \bibinfo {author} {\bibfnamefont {J.~D.}\ \bibnamefont {Joannopoulos}},
  \bibinfo {author} {\bibfnamefont {I.}~\bibnamefont {Celanovic}}, \bibinfo
  {author} {\bibfnamefont {H.}~\bibnamefont {Buljan}}, \ and\ \bibinfo {author}
  {\bibfnamefont {M.}~\bibnamefont {Solja\ifmmode \check{c}\else
  \v{c}\fi{}i\ifmmode~\acute{c}\else \'{c}\fi{}}},\ }\href {\doibase
  10.1103/PhysRevB.85.155422} {\bibfield  {journal} {\bibinfo  {journal} {Phys.
  Rev. B}\ }\textbf {\bibinfo {volume} {85}},\ \bibinfo {pages} {155422}
  (\bibinfo {year} {2012})}\BibitemShut {NoStop}%
\bibitem [{\citenamefont {Svetovoy}, \citenamefont {van Zwol},\ and\
  \citenamefont {Chevrier}(2012)}]{Svetovoy2012prb}%
  \BibitemOpen
  \bibfield  {author} {\bibinfo {author} {\bibfnamefont {V.~B.}\ \bibnamefont
  {Svetovoy}}, \bibinfo {author} {\bibfnamefont {P.~J.}\ \bibnamefont {van
  Zwol}}, \ and\ \bibinfo {author} {\bibfnamefont {J.}~\bibnamefont
  {Chevrier}},\ }\href {\doibase 10.1103/PhysRevB.85.155418} {\bibfield
  {journal} {\bibinfo  {journal} {Phys. Rev. B}\ }\textbf {\bibinfo {volume}
  {85}},\ \bibinfo {pages} {155418} (\bibinfo {year} {2012})}\BibitemShut
  {NoStop}%
\bibitem [{\citenamefont {Wang}\ and\ \citenamefont {Antezza}(2023)}]{JSWM}%
  \BibitemOpen
  \bibfield  {author} {\bibinfo {author} {\bibfnamefont {J.-S.}\ \bibnamefont
  {Wang}}\ and\ \bibinfo {author} {\bibfnamefont {M.}~\bibnamefont {Antezza}},\
  }\href@noop {} {\enquote {\bibinfo {title} {Photon mediated energy, linear
  and angular momentum transport in fullerene and graphene systems beyond local
  equilibrium},}\ } (\bibinfo {year} {2023}),\ \Eprint
  {http://arxiv.org/abs/2307.11361} {arXiv:2307.11361} \BibitemShut {NoStop}%
\bibitem [{\citenamefont {Zhang}, \citenamefont {Antezza},\ and\ \citenamefont
  {Wang}(2022)}]{GrafBilMA}%
  \BibitemOpen
  \bibfield  {author} {\bibinfo {author} {\bibfnamefont {Y.-M.}\ \bibnamefont
  {Zhang}}, \bibinfo {author} {\bibfnamefont {M.}~\bibnamefont {Antezza}}, \
  and\ \bibinfo {author} {\bibfnamefont {J.-S.}\ \bibnamefont {Wang}},\ }\href
  {\doibase https://doi.org/10.1016/j.ijheatmasstransfer.2022.123076}
  {\bibfield  {journal} {\bibinfo  {journal} {Int. J Heat Mass Transf.}\
  }\textbf {\bibinfo {volume} {194}},\ \bibinfo {pages} {123076} (\bibinfo
  {year} {2022})}\BibitemShut {NoStop}%
\bibitem [{\citenamefont {Liu}\ \emph {et~al.}(2022{\natexlab{b}})\citenamefont
  {Liu}, \citenamefont {Ge}, \citenamefont {Yu}, \citenamefont {Cui},\ and\
  \citenamefont {Wu}}]{LiuES2022}%
  \BibitemOpen
  \bibfield  {author} {\bibinfo {author} {\bibfnamefont {R.~Y.}\ \bibnamefont
  {Liu}}, \bibinfo {author} {\bibfnamefont {L.~X.}\ \bibnamefont {Ge}},
  \bibinfo {author} {\bibfnamefont {H.~Y.}\ \bibnamefont {Yu}}, \bibinfo
  {author} {\bibfnamefont {Z.}~\bibnamefont {Cui}}, \ and\ \bibinfo {author}
  {\bibfnamefont {X.~H.}\ \bibnamefont {Wu}},\ }\href {\doibase
  10.30919/es8d529} {\bibfield  {journal} {\bibinfo  {journal} {Eng. Sci.}\
  }\textbf {\bibinfo {volume} {18}},\ \bibinfo {pages} {224} (\bibinfo {year}
  {2022}{\natexlab{b}})}\BibitemShut {NoStop}%
\bibitem [{\citenamefont {Liu}\ and\ \citenamefont {Zhang}(2015)}]{Liu2015APL}%
  \BibitemOpen
  \bibfield  {author} {\bibinfo {author} {\bibfnamefont {X.~L.}\ \bibnamefont
  {Liu}}\ and\ \bibinfo {author} {\bibfnamefont {Z.~M.}\ \bibnamefont
  {Zhang}},\ }\href {\doibase 10.1063/1.4932958} {\bibfield  {journal}
  {\bibinfo  {journal} {Appl. Phys. Lett.}\ }\textbf {\bibinfo {volume}
  {107}},\ \bibinfo {pages} {143114} (\bibinfo {year} {2015})}\BibitemShut
  {NoStop}%
\bibitem [{\citenamefont {Hu}\ \emph {et~al.}(2020)\citenamefont {Hu},
  \citenamefont {Li}, \citenamefont {Zhu},\ and\ \citenamefont
  {Yang}}]{Yang2020prAppl}%
  \BibitemOpen
  \bibfield  {author} {\bibinfo {author} {\bibfnamefont {Y.~Z.}\ \bibnamefont
  {Hu}}, \bibinfo {author} {\bibfnamefont {H.}~\bibnamefont {Li}}, \bibinfo
  {author} {\bibfnamefont {Y.~G.}\ \bibnamefont {Zhu}}, \ and\ \bibinfo
  {author} {\bibfnamefont {Y.}~\bibnamefont {Yang}},\ }\href {\doibase
  10.1103/PhysRevApplied.14.044054} {\bibfield  {journal} {\bibinfo  {journal}
  {Phys. Rev. Appl.}\ }\textbf {\bibinfo {volume} {14}},\ \bibinfo {pages}
  {044054} (\bibinfo {year} {2020})}\BibitemShut {NoStop}%
\bibitem [{\citenamefont {Guizal}\ and\ \citenamefont
  {Felbacq}(1999)}]{Guizal}%
  \BibitemOpen
  \bibfield  {author} {\bibinfo {author} {\bibfnamefont {B.}~\bibnamefont
  {Guizal}}\ and\ \bibinfo {author} {\bibfnamefont {D.}~\bibnamefont
  {Felbacq}},\ }\href {\doibase 10.1016/S0030-4018(99)00192-3} {\bibfield
  {journal} {\bibinfo  {journal} {Optics Communications}\ }\textbf {\bibinfo
  {volume} {165}},\ \bibinfo {pages} {1} (\bibinfo {year} {1999})}\BibitemShut
  {NoStop}%
\bibitem [{\citenamefont {Hwang}(2020)}]{Taiwan_LBF}%
  \BibitemOpen
  \bibfield  {author} {\bibinfo {author} {\bibfnamefont {R.-B.}\ \bibnamefont
  {Hwang}},\ }\href {\doibase 10.1038/s41598-020-69827-w} {\bibfield  {journal}
  {\bibinfo  {journal} {Scientific Reports}\ }\textbf {\bibinfo {volume}
  {10}},\ \bibinfo {pages} {12855} (\bibinfo {year} {2020})}\BibitemShut
  {NoStop}%
\bibitem [{\citenamefont {Jeyar}, \citenamefont {Antezza},\ and\ \citenamefont
  {Guizal}(2023)}]{PRE23}%
  \BibitemOpen
  \bibfield  {author} {\bibinfo {author} {\bibfnamefont {Y.}~\bibnamefont
  {Jeyar}}, \bibinfo {author} {\bibfnamefont {M.}~\bibnamefont {Antezza}}, \
  and\ \bibinfo {author} {\bibfnamefont {B.}~\bibnamefont {Guizal}},\ }\href
  {\doibase 10.1103/PhysRevE.107.025306} {\bibfield  {journal} {\bibinfo
  {journal} {Phys. Rev. E}\ }\textbf {\bibinfo {volume} {107}},\ \bibinfo
  {pages} {025306} (\bibinfo {year} {2023})}\BibitemShut {NoStop}%
\bibitem [{\citenamefont {Jeyar}\ \emph {et~al.}(2023)\citenamefont {Jeyar},
  \citenamefont {Luo}, \citenamefont {Austry}, \citenamefont {Guizal},
  \citenamefont {Zheng}, \citenamefont {Chan},\ and\ \citenamefont
  {Antezza}}]{Luo2023Casimir_gg}%
  \BibitemOpen
  \bibfield  {author} {\bibinfo {author} {\bibfnamefont {Y.}~\bibnamefont
  {Jeyar}}, \bibinfo {author} {\bibfnamefont {M.~G.}\ \bibnamefont {Luo}},
  \bibinfo {author} {\bibfnamefont {K.}~\bibnamefont {Austry}}, \bibinfo
  {author} {\bibfnamefont {B.}~\bibnamefont {Guizal}}, \bibinfo {author}
  {\bibfnamefont {Y.}~\bibnamefont {Zheng}}, \bibinfo {author} {\bibfnamefont
  {H.~B.}\ \bibnamefont {Chan}}, \ and\ \bibinfo {author} {\bibfnamefont
  {M.}~\bibnamefont {Antezza}},\ }\href@noop {} {\enquote {\bibinfo {title}
  {Tunable non-additivity in casimir-lifshitz force between graphene
  gratings},}\ } (\bibinfo {year} {2023}),\ \Eprint
  {http://arxiv.org/abs/2306.17640} {arXiv:2306.17640} \BibitemShut {NoStop}%
\bibitem [{\citenamefont {Zhou}, \citenamefont {Zhang},\ and\ \citenamefont
  {Yi}(2022)}]{Zhou2022langmuir}%
  \BibitemOpen
  \bibfield  {author} {\bibinfo {author} {\bibfnamefont {C.-L.}\ \bibnamefont
  {Zhou}}, \bibinfo {author} {\bibfnamefont {Y.}~\bibnamefont {Zhang}}, \ and\
  \bibinfo {author} {\bibfnamefont {H.-L.}\ \bibnamefont {Yi}},\ }\href
  {\doibase 10.1021/acs.langmuir.2c00467} {\bibfield  {journal} {\bibinfo
  {journal} {Langmuir}\ }\textbf {\bibinfo {volume} {38}},\ \bibinfo {pages}
  {7689} (\bibinfo {year} {2022})}\BibitemShut {NoStop}%
\bibitem [{\citenamefont {Liu}, \citenamefont {Bright},\ and\ \citenamefont
  {Zhang}(2014)}]{Liu2014JHT}%
  \BibitemOpen
  \bibfield  {author} {\bibinfo {author} {\bibfnamefont {X.~L.}\ \bibnamefont
  {Liu}}, \bibinfo {author} {\bibfnamefont {T.~J.}\ \bibnamefont {Bright}}, \
  and\ \bibinfo {author} {\bibfnamefont {Z.~M.}\ \bibnamefont {Zhang}},\ }\href
  {\doibase 10.1115/1.4027802} {\bibfield  {journal} {\bibinfo  {journal} {J.
  Heat Transfer}\ }\textbf {\bibinfo {volume} {136}},\ \bibinfo {pages}
  {092703} (\bibinfo {year} {2014})}\BibitemShut {NoStop}%
\bibitem [{\citenamefont {Biehs}\ \emph {et~al.}(2011)\citenamefont {Biehs},
  \citenamefont {Ben-Abdallah}, \citenamefont {Rosa}, \citenamefont {Joulain},\
  and\ \citenamefont {Greffet}}]{Biehs2011OE}%
  \BibitemOpen
  \bibfield  {author} {\bibinfo {author} {\bibfnamefont {S.-A.}\ \bibnamefont
  {Biehs}}, \bibinfo {author} {\bibfnamefont {P.}~\bibnamefont {Ben-Abdallah}},
  \bibinfo {author} {\bibfnamefont {F.~S.~S.}\ \bibnamefont {Rosa}}, \bibinfo
  {author} {\bibfnamefont {K.}~\bibnamefont {Joulain}}, \ and\ \bibinfo
  {author} {\bibfnamefont {J.-J.}\ \bibnamefont {Greffet}},\ }\href {\doibase
  10.1364/OE.19.0A1088} {\bibfield  {journal} {\bibinfo  {journal} {Opt.
  Express}\ }\textbf {\bibinfo {volume} {19}},\ \bibinfo {pages} {A1088}
  (\bibinfo {year} {2011})}\BibitemShut {NoStop}%
\bibitem [{\citenamefont {Messina}\ and\ \citenamefont
  {Antezza}(2014)}]{Messina2014}%
  \BibitemOpen
  \bibfield  {author} {\bibinfo {author} {\bibfnamefont {R.}~\bibnamefont
  {Messina}}\ and\ \bibinfo {author} {\bibfnamefont {M.}~\bibnamefont
  {Antezza}},\ }\href {\doibase 10.1103/PhysRevA.89.052104} {\bibfield
  {journal} {\bibinfo  {journal} {Phys. Rev. A}\ }\textbf {\bibinfo {volume}
  {89}},\ \bibinfo {pages} {052104} (\bibinfo {year} {2014})}\BibitemShut
  {NoStop}%
\bibitem [{\citenamefont {Messina}\ and\ \citenamefont
  {Antezza}(2011)}]{Messina2011PRA}%
  \BibitemOpen
  \bibfield  {author} {\bibinfo {author} {\bibfnamefont {R.}~\bibnamefont
  {Messina}}\ and\ \bibinfo {author} {\bibfnamefont {M.}~\bibnamefont
  {Antezza}},\ }\href {\doibase 10.1103/PhysRevA.84.042102} {\bibfield
  {journal} {\bibinfo  {journal} {Phys. Rev. A}\ }\textbf {\bibinfo {volume}
  {84}},\ \bibinfo {pages} {042102} (\bibinfo {year} {2011})}\BibitemShut
  {NoStop}%
\bibitem [{\citenamefont {Messina}\ \emph {et~al.}(2017)\citenamefont
  {Messina}, \citenamefont {Noto}, \citenamefont {Guizal},\ and\ \citenamefont
  {Antezza}}]{Messina2017PRB}%
  \BibitemOpen
  \bibfield  {author} {\bibinfo {author} {\bibfnamefont {R.}~\bibnamefont
  {Messina}}, \bibinfo {author} {\bibfnamefont {A.}~\bibnamefont {Noto}},
  \bibinfo {author} {\bibfnamefont {B.}~\bibnamefont {Guizal}}, \ and\ \bibinfo
  {author} {\bibfnamefont {M.}~\bibnamefont {Antezza}},\ }\href {\doibase
  10.1103/PhysRevB.95.125404} {\bibfield  {journal} {\bibinfo  {journal} {Phys.
  Rev. B}\ }\textbf {\bibinfo {volume} {95}},\ \bibinfo {pages} {125404}
  (\bibinfo {year} {2017})}\BibitemShut {NoStop}%
\bibitem [{\citenamefont {Palik}(1998)}]{Palik}%
  \BibitemOpen
  \bibfield  {author} {\bibinfo {author} {\bibfnamefont {E.}~\bibnamefont
  {Palik}},\ }\href@noop {} {\emph {\bibinfo {title} {Handbook of Optical
  Constants of Solids}}}\ (\bibinfo  {publisher} {Academic},\ \bibinfo
  {address} {New York},\ \bibinfo {year} {1998})\BibitemShut {NoStop}%
\bibitem [{\citenamefont {Falkovsky}\ and\ \citenamefont
  {Varlamov}(2007)}]{Falkovsky2007}%
  \BibitemOpen
  \bibfield  {author} {\bibinfo {author} {\bibfnamefont {L.~A.}\ \bibnamefont
  {Falkovsky}}\ and\ \bibinfo {author} {\bibfnamefont {A.~A.}\ \bibnamefont
  {Varlamov}},\ }\href {\doibase 10.1140/epjb/e2007-00142-3} {\bibfield
  {journal} {\bibinfo  {journal} {Eur. Phys. J. B}\ }\textbf {\bibinfo {volume}
  {56}},\ \bibinfo {pages} {281} (\bibinfo {year} {2007})}\BibitemShut
  {NoStop}%
\bibitem [{\citenamefont {Falkovsky}(2008)}]{Falkovsky2008}%
  \BibitemOpen
  \bibfield  {author} {\bibinfo {author} {\bibfnamefont {L.~A.}\ \bibnamefont
  {Falkovsky}},\ }\href {\doibase 10.1088/1742-6596/129/1/012004} {\bibfield
  {journal} {\bibinfo  {journal} {J. Phys. Conf. Ser.}\ }\textbf {\bibinfo
  {volume} {129}},\ \bibinfo {pages} {012004} (\bibinfo {year}
  {2008})}\BibitemShut {NoStop}%
\bibitem [{\citenamefont {Awan}\ \emph {et~al.}(2016)\citenamefont {Awan},
  \citenamefont {Lombardo}, \citenamefont {Colli}, \citenamefont {Privitera},
  \citenamefont {Kulmala}, \citenamefont {Kivioja}, \citenamefont {Koshino},\
  and\ \citenamefont {Ferrari}}]{Awan_2016}%
  \BibitemOpen
  \bibfield  {author} {\bibinfo {author} {\bibfnamefont {S.~A.}\ \bibnamefont
  {Awan}}, \bibinfo {author} {\bibfnamefont {A.}~\bibnamefont {Lombardo}},
  \bibinfo {author} {\bibfnamefont {A.}~\bibnamefont {Colli}}, \bibinfo
  {author} {\bibfnamefont {G.}~\bibnamefont {Privitera}}, \bibinfo {author}
  {\bibfnamefont {T.~S.}\ \bibnamefont {Kulmala}}, \bibinfo {author}
  {\bibfnamefont {J.~M.}\ \bibnamefont {Kivioja}}, \bibinfo {author}
  {\bibfnamefont {M.}~\bibnamefont {Koshino}}, \ and\ \bibinfo {author}
  {\bibfnamefont {A.~C.}\ \bibnamefont {Ferrari}},\ }\href {\doibase
  10.1088/2053-1583/3/1/015010} {\bibfield  {journal} {\bibinfo  {journal} {2D
  Mater.}\ }\textbf {\bibinfo {volume} {3}},\ \bibinfo {pages} {015010}
  (\bibinfo {year} {2016})}\BibitemShut {NoStop}%
\bibitem [{\citenamefont {Gu\'erout}\ \emph {et~al.}(2012)\citenamefont
  {Gu\'erout}, \citenamefont {Lussange}, \citenamefont {Rosa}, \citenamefont
  {Hugonin}, \citenamefont {Dalvit}, \citenamefont {Greffet}, \citenamefont
  {Lambrecht},\ and\ \citenamefont {Reynaud}}]{Greffet2012prb_trans_def}%
  \BibitemOpen
  \bibfield  {author} {\bibinfo {author} {\bibfnamefont {R.}~\bibnamefont
  {Gu\'erout}}, \bibinfo {author} {\bibfnamefont {J.}~\bibnamefont {Lussange}},
  \bibinfo {author} {\bibfnamefont {F.~S.~S.}\ \bibnamefont {Rosa}}, \bibinfo
  {author} {\bibfnamefont {J.-P.}\ \bibnamefont {Hugonin}}, \bibinfo {author}
  {\bibfnamefont {D.~A.~R.}\ \bibnamefont {Dalvit}}, \bibinfo {author}
  {\bibfnamefont {J.-J.}\ \bibnamefont {Greffet}}, \bibinfo {author}
  {\bibfnamefont {A.}~\bibnamefont {Lambrecht}}, \ and\ \bibinfo {author}
  {\bibfnamefont {S.}~\bibnamefont {Reynaud}},\ }\href {\doibase
  10.1103/PhysRevB.85.180301} {\bibfield  {journal} {\bibinfo  {journal} {Phys.
  Rev. B}\ }\textbf {\bibinfo {volume} {85}},\ \bibinfo {pages} {180301}
  (\bibinfo {year} {2012})}\BibitemShut {NoStop}%
\bibitem [{\citenamefont {Liu}\ and\ \citenamefont
  {Zhang}(2014)}]{Liu2014apl_trans_def}%
  \BibitemOpen
  \bibfield  {author} {\bibinfo {author} {\bibfnamefont {X.~L.}\ \bibnamefont
  {Liu}}\ and\ \bibinfo {author} {\bibfnamefont {Z.~M.}\ \bibnamefont
  {Zhang}},\ }\href {\doibase 10.1063/1.4885396} {\bibfield  {journal}
  {\bibinfo  {journal} {Appl. Phys. Lett.}\ }\textbf {\bibinfo {volume}
  {104}},\ \bibinfo {pages} {251911} (\bibinfo {year} {2014})}\BibitemShut
  {NoStop}%
\bibitem [{\citenamefont {Zhou}\ \emph {et~al.}(2022)\citenamefont {Zhou},
  \citenamefont {Zhang}, \citenamefont {Torbatian}, \citenamefont {Novko},
  \citenamefont {Antezza},\ and\ \citenamefont {Yi}}]{Zhou2022prm}%
  \BibitemOpen
  \bibfield  {author} {\bibinfo {author} {\bibfnamefont {C.-L.}\ \bibnamefont
  {Zhou}}, \bibinfo {author} {\bibfnamefont {Y.}~\bibnamefont {Zhang}},
  \bibinfo {author} {\bibfnamefont {Z.}~\bibnamefont {Torbatian}}, \bibinfo
  {author} {\bibfnamefont {D.}~\bibnamefont {Novko}}, \bibinfo {author}
  {\bibfnamefont {M.}~\bibnamefont {Antezza}}, \ and\ \bibinfo {author}
  {\bibfnamefont {H.-L.}\ \bibnamefont {Yi}},\ }\href {\doibase
  10.1103/PhysRevMaterials.6.075201} {\bibfield  {journal} {\bibinfo  {journal}
  {Phys. Rev. Mater.}\ }\textbf {\bibinfo {volume} {6}},\ \bibinfo {pages}
  {075201} (\bibinfo {year} {2022})}\BibitemShut {NoStop}%
\end{thebibliography}%

\end{document}